\documentclass[aps,prd,twocolumn,superscriptaddress,groupedaddress,preprintnumbers,showkeys,showpacs,nofootinbib]{revtex4}  
\usepackage[latin1]{inputenc}                    % Codepage latin1
\usepackage{graphicx}  % needed for figures
\usepackage{dcolumn}   % needed for some tables
\usepackage{bm}        % for math
\usepackage{textcomp}
\usepackage{epsfig}
\usepackage{epsf}
\usepackage{bbm}
\usepackage{latexsym}                            % Load additional symbols
\usepackage{amsfonts}                            % Use AMS fonts,
\usepackage{amssymb}                             % symbols,
\usepackage{amsmath}                             % and definitions
\usepackage[mathscr]{eucal}                      % Euler font symbols
\usepackage{dcolumn}                             % Align table columns
\usepackage{multirow}                               % on decimal point
\usepackage{bm}                                  % Bold math
\usepackage{hyperref}                            % References as links
\usepackage[usenames]{color}
\usepackage{array}
\usepackage{appendix}
\usepackage{bbold}
\usepackage{soul,xcolor}
\setstcolor{red}
\graphicspath{{.}{Graphics/}}                    % Path for graphics
\newcommand{\be}{\begin{equation}}
\newcommand{\ee}{\end{equation}}
\newcommand{\bea}{\begin{eqnarray}}
\newcommand{\eea}{\end{eqnarray}}

\def \3{\ss }

\newcommand{\beq}{\begin{eqnarray}}
\newcommand{\eeq}{\end{eqnarray}}

\renewcommand{\arraystretch}{1.8}

\usepackage{ulem}
\usepackage{letltxmacro}

\newif\ifcorrectingmode
\correctingmodetrue
\begin{document}

\title{Nucleon axial, tensor, and scalar charges and $\sigma$-terms in lattice QCD}

\author{C.~Alexandrou} \affiliation{Department of Physics, University of Cyprus, P.O. Box 20537, 1678 Nicosia, Cyprus} \affiliation{Computation-based Science and Technology Research Center, The Cyprus Institute, 20 Kavafi Str., Nicosia 2121, Cyprus}
\author{S.~Bacchio} \affiliation{Computation-based Science and Technology Research Center, The Cyprus Institute, 20 Kavafi Str., Nicosia 2121, Cyprus}
\author{M.~Constantinou} \affiliation{Department of Physics, Temple University, 1925 N. 12th Street, Philadelphia, PA 19122-1801, USA}
\author{J.~Finkenrath} \affiliation{Computation-based Science and Technology Research Center, The Cyprus Institute, 20 Kavafi Str., Nicosia 2121, Cyprus} 
\author{K.~Hadjiyiannakou} \affiliation{Computation-based Science and Technology Research Center, The Cyprus Institute, 20 Kavafi Str., Nicosia 2121, Cyprus}
\author{K.~Jansen} \affiliation{NIC, DESY, Platanenallee 6, D-15738 Zeuthen, Germany}
\author{G.~Koutsou} \affiliation{Computation-based Science and Technology Research Center, The Cyprus Institute, 20 Kavafi Str., Nicosia 2121, Cyprus}
\author{A.~Vaquero Aviles-Casco} \affiliation{Department of Physics and Astronomy, University of Utah, Salt Lake City, UT 84112, USA\\[0.2cm]{\centering	%\includegraphics[width=0.18\linewidth]{logo2}\\[0.2cm]
}}
%
%\collaboration{Extended Twisted Mass Collaboration} \noaffiliation

%\date{\today}
\begin{abstract}

 We  determine  the nucleon axial, scalar and tensor charges   within  lattice  Quantum Chromodynamics including all  contributions from valence and sea quarks. 
 We analyze three gauge ensembles simulated within the twisted mass formulation at approximately physical value of the pion mass. Two of these ensembles are simulated with two dynamical light quarks and lattice spacing $a=0.094$~fm and the third with $a=0.08$~fm includes in addition the strange and charm quarks in the sea. After comparing the results among these three ensembles, we quote as final values our most accurate analysis using the latter ensemble.
 For the nucleon isovector axial charge  we find  $1.286(23)$ in agreement with the experimental value. We provide the flavor decomposition of the intrinsic spin $\frac{1}{2}\Delta\Sigma^q$ carried by quarks in the nucleon obtaining for the up, down, strange and charm quarks  $\frac{1}{2}\Delta\Sigma^{u}=0.431(8)$, $\frac{1}{2}\Delta\Sigma^{d}=-0.212(8)$, $\frac{1}{2}\Delta\Sigma^{s}=-0.023(4)$ and $\frac{1}{2}\Delta\Sigma^{c}=-0.005(2)$, respectively. The corresponding values of the tensor and scalar charges for each quark flavor are also evaluated providing valuable input for experimental searches for beyond the standard model physics.
 In addition, we extract the nucleon $\sigma$-terms and find for the light quark content $\sigma_{\pi N}=41.6(3.8)$~MeV and  for the strange $\sigma_{s}=45.6(6.2)$~MeV. The y-parameter that is used in phenomenological studies  we find  $y=0.078(7)$.
       \end{abstract}

\maketitle
\section{Introduction}
The nucleon axial charge, denoted here by $g_A^{u-d}$, is a fundamental quantity within the Standard Model (SM) of particle physics.  It determines the rate of the weak decay of neutrons into protons and provides a 
quantitative measure of spontaneous chiral symmetry breaking in hadronic physics. 
It enters in the analysis of neutrinoless double-beta decay and in the unitarity 
tests of the Cabibbo-Kobayashi-Maskawa matrix. It is known precisely from neutron beta decay measurements 
using polarized ultracold neutrons~\cite{Mendenhall:2012tz,Mund:2012fq,Markisch:2018ndu}. Partial conservation of the axial current (PCAC) relates the axial and pseudoscalar charges and allows us to predict the latter. The  flavor-diagonal axial charge $g_A^f$ determines the intrinsic spin $\frac{1}{2}\Delta\Sigma^q$ carried by the quarks in the nucleon. These are being measured in deep inelastic scattering (DIS) experiments in major facilities such as Jefferson lab and CERN and are targeted in the program of the Electron Ion Collider (EIC). 

The isovector tensor and scalar charges can put limits on the existence of beyond SM interactions with  scalar and tensor structures~\cite{Bhattacharya:2011qm}. Ongoing  neutrino scattering experiments  probing scalar and/or tensor interactions include the experiments DUNE~\cite{PhysRevD.99.036006},  COHERENT~\cite{Akimov1123}, IsoDAR~\cite{Abs:2015tbh}, LZ~\cite{Akerib:2015cja},  GEMMA~\cite{Beda:2012zz} and  the TEXONO collaboration~\cite{PhysRevD.75.012001}.
A review on probing new physics by CP violating processes using the electric dipole moments of atoms can be found in Ref.~\cite{refId0}. High precision measurements of spectral lines in few-electron atoms can  probe the existence of exotic forces between electrons~\cite{PhysRevD.96.115002}, while
Direct dark matter searches look for new scalar interactions~\cite{Undagoitia:2015gya} and semileptonic kaon~\cite{CHIZHOV1996359}  or tau~\cite{Dhargyal:2016jgo} decays experiments are probing for tensor interactions. The tensor and scalar charges
 are less precisely known, and a determination within lattice QCD can provide essential input for precision measurements  probing  the existence of  novel scalar and tensor interactions aiding experimental searches. The tensor charge is the first Mellin moment of the transversity parton  distribution function (PDF) being  studied in many experiments including Drell-Yan and semi-inclusive DIS by COMPASS at CERN~\cite{Adolph:2014fjw} and at Jefferson Lab. The planned SoLID experiment at Jefferson Lab~\cite{Ye:2016prn} will allow to measure the tensor charge  with an improved accuracy. The extraction of the transversity distribution is less precise than
the unpolarized PDF, and 
additional phenomenological modeling is required. In addition, the  flavor-diagonal tensor charge enters into the determination of the quark electric dipole moment contribution to the
neutron electric dipole moment~\cite{Bhattacharya:2016zcn}, which  signals CP violation.

The nucleon matrix element of the single-flavor scalar operator  $g_S^f$ is directly connected to the quark  content of the nucleon, or the so-called nucleon $\sigma^f$-term, which determines the mass generated
by a quark in the nucleon and it is, thus, related to the explicit breaking of chiral symmetry~\cite{glashow1968breaking}. Nucleon $\sigma$-terms are relevant for pion and kaon nucleon scattering processes but also for the interpretation of direct-detection dark matter searches. The dark matter candidates under consideration are weakly interacting massive particles  in a number of beyond the SM  theories that interact with normal matter by elastic scattering with nuclei. Besides its direct relation to the $\sigma$-term, the isovector scalar charge $g_S^{u-d}$ measures the proportionality constant between  the neutron-proton mass splitting $\delta m_N^{\rm QCD}$ and the up and down quark mass splitting $\delta m_{ud}$  in the absence of electromagnetism via the relation $\delta m_N^{\rm QCD}=g_S^{u-d}\delta m_{ud}$~\cite{Gonzalez-Alonso:2013ura}. This relation first appeared in Ref.~\cite{Gasser:1982ap}, while a similar result was derived   in Ref.~\cite{Crewther:1979pi}.
The fundamental role of these quantities
in the physics of weak 
interactions and in beyond the SM physics makes their
non-perturbative determination of central importance.

The non-perturbative nature of the fundamental theory of the strong interaction  makes a theoretical calculation of these fundamental quantities  difficult. The discretized version  of the theory defined on a four-dimensional Euclidean lattice and known as lattice Quantum Chromodynamics (QCD) provides a
rigorous, non-perturbative formulation  that  
allows for a numerical simulation  with  controlled systematic uncertainties.
Since, as mentioned already, $g_A^{u-d}$ is accurately measured experimentally it serves as a benchmark quantity for lattice QCD. Numerous past lattice QCD  studies~\cite{Lin:2017snn}
underestimated $g_A^{u-d}$ and impeded 
reliable predictions of the other nucleon charges.     
It is only recently that an accurate computation of $g_A^{u-d}$ was presented~\cite{Chang:2018uxx} 
that reproduced the experimental value. It was, however, obtained using chiral extrapolations involving  ensembles with heavier than physical pions.  For a complete list of lattice QCD results with details on the lattice QCD framework used, we refer to the recent FLAG report~\cite{Aoki:2019cca}.
 Reproducing the value of $g_A^{u-d}$ within a lattice QCD framework serves both as a validation and as a 
most valuable benchmark computation for the extraction of  the isovector scalar $g_S^{u-d}$  and tensor $g_T^{u-d}$ charges. In addition, a precise computation of $g_A^{u-d}$ in lattice QCD can provide a constraint for non-standard right-handed currents~\cite{Bhattacharya:2011qm}.

In this work, we compute the nucleon charges and $\sigma$-terms
 using  gauge configurations generated with the physical values of the light quark masses, avoiding chiral extrapolation or any 
modeling of the pion mass dependence. We consider  two ensembles with two light quarks in the sea, denoted by $N_f=2$ ensembles, and one ensemble where, besides the light quarks, we include the strange and charm quarks in the sea, denoted with $N_f=2+1+1$.  The latter ensemble provides one of the best description of the QCD vacuum to date and thus we devote most of our computational resources to its analysis and use it to extract our final values.
For this $N_f=2+1+1$ ensemble we  achieve high precision not only for the isovector axial (A), tensor (T) and scalar (S) quantities  but also for the single flavor charges and $\sigma$-terms.
Such an accurate computation from first principles of the axial, scalar and tensor charges  
for each quark flavor, 
as well as the direct determination of the $\pi N$, strange and charm $\sigma$-terms,  constitutes a major step in our understanding of the structure of the nucleon.

The remainder of this paper is organized as follows: in Section~\ref{sec:methodology} we provide the methodology used for extracting the nucleon charges using lattice QCD, in Section~\ref{sec:analysis} we detail the analysis carried out, in particular as regards ensuring suppression of excited states, and provide unremormalized results of the nucleon charges. In Section~\ref{sec:renormalization} we describe our renormalization procedure and in Section~\ref{sec:results} we provide renormalized results for the nucleon charges and compare with phenomenology and other lattice results. In Section~\ref{sec:conclusions} we review our final results and provide our conclusions.

\section{Methodology}
\label{sec:methodology}
The axial, tensor and scalar flavor charges $g^f_{\rm A,T,S}$  are obtained from  the nucleon matrix elements of the axial, tensor and scalar operators at zero momentum transfer, given by
\begin{equation}\label{eq:charges}
\langle N|\bar{\psi}^f\Gamma_{\rm A,S,T}\psi^f|N\rangle = g^f_{\rm A,T,S} \bar{u}_N\Gamma_{\rm A,S,T} u _N\,,
\end{equation}
where $u_N$ is  the nucleon spinor, $f$ denotes the quark flavor, and $\Gamma_A=\gamma_\mu\gamma_5$ for the axial-vector operator, $\Gamma_S=\mathbb{1}$ for the scalar  and $\Gamma_T=\sigma_{\mu\nu}$ for the tensor. The renormalization group invariant
$\sigma^f$-term is defined by $m_f\langle N|\bar{\psi}_f\psi_f|N\rangle$ where $m_f$ is the quark mass.

\subsection{Lattice QCD formulation and gauge ensembles}
We use three gauge ensembles  simulated   with a physical value of the pion mass~\cite{Abdel-Rehim:2015pwa,Alexandrou:2018egz} using the twisted mass fermion discretization scheme~\cite{Frezzotti:2000nk,Frezzotti:2003ni} with a clover-term~\cite{Sheikholeslami:1985ij}.  The parameters are listed in Table~\ref{tab:params}. We refer to these ensembles as physical point ensembles.  
Twisted mass fermions (TMF) provide an attractive formulation for lattice QCD allowing for automatic ${\cal O}(a)$ improvement~\cite{Frezzotti:2003ni}, where $a$ is the lattice spacing. This is an important property for evaluating the quantities  considered here, since all quantities have lattice artifacts of ${\cal O}(a^2)$ and are closer to the continuum limit as observed in previous studies using simulation with larger than physical  pion mass~\cite{Alexandrou:2010hf}. A  clover-term is added to the TMF action to allow for smaller $\mathcal{O}(a^2)$ breaking effects between the neutral and charged pions that lead to the stabilization of simulations with light quark masses close to the  physical pion mass. For more details on the TMF formulation  see Refs.~\cite{Frezzotti:2005gi,Boucaud:2008xu} and for the simulation strategy Refs.~\cite{Abdel-Rehim:2015pwa,Alexandrou:2018egz}.
\begin{table}
	\begin{flushleft}
	%\resizebox{1.\linewidth}{!}{
	\small
		\renewcommand{\arraystretch}{1.2}
		\renewcommand{\tabcolsep}{2.5pt}
		\begin{tabular}{ c c c  c   c c }
			\toprule
			Ensemble     & $L^3\times T$ & $m_N/m_\pi$    & $m_\pi L$ & $m_\pi$ [MeV] & L [fm]   \\ \hline
			\multicolumn{6}{c}{$N_f=2$, $\beta=2.1$, $a=0.0938(3)(1)$~fm}\\
			% $a$ [fm]     & $m_\pi$ [MeV]   &  \\ \hline\hline
			cA2.09.48     & $48^3 \times 98$ &7.15(2)       & 2.98    &   130.3(4)(2)& 4.50(1)         \\
			% 0.0938(3)(1) & 130.3(4)(2)   &     &  \\ \hline
			cA2.09.64     & $64^3 \times 128$ & 7.14(4) & 3.97       & 130.6(4)(2)  & 6.00(2)      \\
			% 0.0938(3)(1) & 130.6(4)(2)   & 7.15(2)       &  \\ \hline
			\hline
			\multicolumn{6}{c}{$N_f=2+1+1$, $\beta=1.778$, $a=0.0801(4)$~fm}\\
			cB211.072.64  & $64^3 \times 128$ & 6.74(3) &  3.62 & 139.3(7)  &  5.12(3)     \\
			% 0.0801(4)    & 139.3(7)      & 6.74(3)       &  \\ \hline
			\botrule
		\end{tabular}
	%}
	\end{flushleft}
	\vspace*{-0.4cm}
	\caption{Twisted mass fermion ensembles simulated with  clover improvement at the physical pion mass~\cite{Abdel-Rehim:2015pwa,Alexandrou:2018egz}.  $N_f$ is the number of quark flavors in the sea,  $L$ ($T$) is the spatial (temporal)  extent of the lattice in lattice units and $a$ is the lattice spacing determined using the nucleon mass. When two errors are given, the first is statistical and the second is systematic. }
	%The first parenthesis gives the statistical error and the second the systematic taken as the difference in the value of $a$ determined from the pion mass and from gluonic quantities.
	\label{tab:params}
\end{table}

The two ensembles denoted by cA2.09.48 and cA2.09.64, are generated  with two dynamical  mass degenerate up and down quarks ($N_f=2$) with mass tuned to reproduce the physical pion mass~\cite{Abdel-Rehim:2015pwa}. They have the same lattice spacing but
use two lattice sizes of  $48^3\,\textsf{x}\,96$ and $64^3\,\textsf{x}\, 128$ allowing for checking finite volume dependence. The ensemble denoted by cB211.072.64  has been generated on a lattice of size $64^3\,\textsf{x}\, 128$ with two degenerate light quarks and the strange and charm quarks ($N_f=2$+1+1) in the sea with masses tuned to produce the physical pion, kaon and $D_s$-meson mass, respectively, keeping the ratio of charm to strange quark mass $m_c/m_s\simeq11.8$~\cite{Aoki:2019cca}.
For the  valence strange and charm quarks we use Osterwalder-Seiler fermions~\cite{Osterwalder:1977pc} with mass tuned to reproduce the $\Omega^{-}$ and the $\Lambda^{+}_c$ baryons~\cite{Alexandrou:2017xwd}, respectively. 
Results for nucleon charges using the cA2.09.48 ensemble have been presented in Refs.~\cite{Alexandrou:2017hac, Alexandrou:2017qyt, Alexandrou:2017oeh}. Since we perform a reanalysis to match our analysis strategy for the cB211.072.64 ensemble, the results are updated.
%\medskip

\subsection{Computation of correlators}
The nucleon matrix elements are extracted
by computing appropriately defined three-point correlators $C^{f}_{\rm A,S,T}$, as well as  the nucleon  two-point correlators, $C_{\rm 2pt}$, at zero momentum. These  correlation functions are constructed by creating a state from the vacuum with the quantum numbers of  the nucleon  at some initial time  (source)   that is  annihilated  at a later time $t_s$ (sink), where we take the source time to be zero.
All expressions that follow are given in Euclidean space.
We consider  three-point correlators 
\begin{align}
&C^{f}_{\rm A,S,T}(P;t_s,t_{\rm ins}) {=}\nonumber\\ 
&\sum_{\vec{x}_{\rm ins},\vec{x}_s} 
\textrm{Tr} \left[P \langle J_N(t_s,\vec{x}_s) \mathcal{O}_{\rm A,S,T}^{f}(t_{\rm ins},\vec{x}_{\rm ins}) \bar{J}_N(0,\vec{0}) \rangle \right],\label{eq:thrp}
\end{align}
where $\mathcal{O}_{\rm A,S,T}^{f}(t_{\rm ins},\vec{x}_{\rm ins})$
is a local current operator that couples to a quark at insertion time $t_{\rm ins}$ having
$0 \leq t_{\rm ins} \leq t_s$.
 $P$ is a
projector acting on spin indices, and we will use either the so-called
unpolarized projector $P_0 {=} \frac{1}{2}(1{+}\gamma_0)$ or the three
polarized $P_k{=}i \gamma_5 \gamma_k P_0 $ combinations. For
$J_N$, we use the standard nucleon interpolating operator,
\begin{equation}
J_N(x)=\epsilon^{abc}u^a(x)[u^{\intercal b}(x)\mathcal{C}\gamma_5d^c(x)]\,,
\end{equation}
where $u$ and $d$ are up- and down-quark spinors and
$\mathcal{C}{=}\gamma_0 \gamma_2$ is the charge conjugation
matrix. The local current operator is given by
\begin{equation}
\mathcal{O}_{\rm A,S,T}^{f}(x) = \bar{\psi}^f(x)\,\Gamma_{\rm A,S,T}\,\psi^f(x)
\end{equation}
where $\psi^f$ is a quark spinor of flavor $f$ and the matrices $\Gamma_{\rm A,S,T}$ are defined in Eq.~\eqref{eq:charges}.

Inserting two complete sets of states in Eq.~(\ref{eq:thrp}), one
obtains a tower of hadron matrix elements with the quantum numbers of
the nucleon multiplied by overlap terms and time dependent
exponentials. For large enough time separations, the excited state
contributions are suppressed compared to the nucleon ground state and
one can then extract the desired matrix element. Knowledge of
two-point functions is required in order to cancel time dependent
exponentials and overlaps. They are given by
\begin{equation}
C_{\rm 2pt}(t_s) {=} 
\sum_{\vec{x}_s} 
\textrm{Tr}  \left[ P_0 
{\langle}J_N(t_s,\vec{x}_s) \bar{J}_N(0,\vec{0}) {\rangle}
\right].
\label{eq:twop}
\end{equation}

In order to increase the overlap of the interpolating operator $J_N$
with the nucleon state and thus decrease overlap with excited states we
use Gaussian smeared quark
fields via~\cite{Alexandrou:1992ti,Gusken:1989qx}:
\begin{align}
\psi_{\rm smear}^a(t,{\vec{x}}) &= \sum_{\vec{y}} F^{ab}({\vec{x}},{\vec{y}};U(t))\ \psi^b(t,{\vec{y}})\,,\\  
F &= (\mathbbm{1} + {\alpha} H)^{n} \,, \nonumber\\  
H({\vec{x}},{\vec{y}}; U(t)) &= \sum_{i=1}^3[U_i(x) \delta_{x,y-\hat\imath} + U_i^\dagger(x-\hat\imath) \delta_{x,y+\hat\imath}],
\end{align}
with APE-smearing~\cite{Albanese:1987ds} applied to the gauge fields
$U_\mu$ entering the Gaussian smearing hopping matrix $H$. For the APE
smearing~\cite{Albanese:1987ds} we use 50 iteration steps and
$\alpha_\textrm{APE}{=}0.5$. The Gaussian smearing parameters are
tuned to yield approximately a root mean square radius for the nucleon
of about 0.5~fm, which has been found to yield early convergence to 
the nucleon two-point functions. This can be
achieved by a combination of the smearing parameters $\alpha$ and
$n$. We use $\alpha$=0.2, 0.2, and 4.0 and $n$=125, 90, and 50 for
ensembles cB211.072.64, cA2.09.64, and cA2.09.48, respectively.
We employ a multi-grid solver~\cite{Frommer:2013fsa} to speed up the inversions that has been extended to the case of  the twisted mass operator and shown to yield a speed-up
of more than one order of magnitude at the
physical point compared to the conjugate gradient method (CG)~\cite{Alexandrou:2016izb}. The resulting propagators are used to construct two- and three-point correlators.
%\medskip

\subsection{Connected and disconnected contributions}
The three-point correlators receive two contributions, one arising when the current couples to a valence quark and one when coupled to a sea quark. The former is referred to as giving rise to a connected and the latter to a  disconnected contribution. The connected contributions are evaluated using  sequential inversions through the sink. Since in this method $t_s$ and the four spin projection matrices needed for the extraction of the charges are fixed, four sets of sequential inversions are performed for each value of $t_s$ in the rest frame of the nucleon.  In Table~\ref{tab:statistics} we give the statistics used for computing the connected contributions for the three ensembles analyzed. As can be seen, the statistics are increased as we increase $t_s$   to keep statistical errors comparable for all time separations $t_s$. Ensemble cB211.072.64 has the largest statistics and we will thus base our final values on this ensembles.
\begin{table}
	\begin{flushleft}
		{\small
			\renewcommand{\arraystretch}{1.2}
			\renewcommand{\tabcolsep}{3pt}
			\begin{tabular}{l|rrrrrrr}
				\toprule
				\multicolumn{1}{r|}{$t_s/a$} &   8 &   10 &   12 &   14 &    16 &    18 &    20 \\ \hline
				\multicolumn{1}{r|}{$t_s$~[fm]} & 0.75 & 0.94 & 1.13 & 1.31 & 1.50 & 1.69 & 1.88\\
				cA2.09.48                    &  -- & 9264 & 9264 & 9264 & 47696 & 69784 &    -- \\
				cA2.09.64                    &  -- &   -- & 5328 & 8064 & 17008 &    -- &    -- \\ \hline
				\multicolumn{1}{r|}{$t_s$~[fm]} & 0.64 & 0.80 & 0.96 & 1.12 & 1.28 & 1.44 & 1.60\\
				cB211.072.64                 & 750 & 1500 & 3000 & 4500 & 12000 & 36000 & 48000 \\
			\botrule              
			\end{tabular}
		}
	\end{flushleft}
	\vspace*{-0.4cm}
	\caption{The values of the  sink-source time separation in lattice units, $t_s/a$, and in physical units, $t_s$, and the associated statistics used for the computation of the connected contribution to the three-point function for the three  ensembles listed in Table~\ref{tab:params}. For the ensemble cA2.09.48, $t_s/a=16$ and $t_a/a=18$ are computed only for the case of the  scalar charge.}
	\label{tab:statistics}
\end{table}
For the  disconnected contributions  we  utilize a combination of methods that are suitable for physical point ensembles~\cite{Alexandrou:2018sjm}. These employ full dilution in spin and color in order to eliminate exactly any contamination from off-diagonal elements and a  partial dilution in space-time using Hierarchical Probing~\cite{Stathopoulos:2013aci} up to a distance of $2^3$ lattice units taking advantage of the exponential decay of off-diagonal elements with the distance. For the up and down quarks  this exponential decay is slow
and therefore we  combine  with deflation of the low modes. For the strange and charm quarks deflation is not necessary due to the heavier mass and we use a distance of $2^3$ and $2^2$ lattice units, respectively, in the Hierarchical Probing. Additionally, we  employ the so-called one-end trick~\cite{Michael:2007vn,McNeile:2006bz} that makes use of the properties of the twisted mass action to improve the signal-to-noise ratio.
Deflation of low modes and Hierarchical Probing have been employed only for the computations using the cB211.072.64 ensemble. For the cA2.09.48 ensemble we use stochastic sources.
In Table~\ref{tab:disconnected} we list the  parameters and statistics   used in the calculation of the disconnected contributions. 
%\medskip
\begin{table}
		{\small
			\renewcommand{\arraystretch}{1.2}
			\renewcommand{\tabcolsep}{3pt}
			\begin{tabular}{l|ccccccc}
				\toprule
				 &$C_{2pt}$ & $C^{u+d}_{\rm loop}$& $C^{s}_{\rm loop}$& $C^{c}_{\rm loop}$\\\hline
				cA2.09.48 & 848000 & 4808250 & 2204672&2691250\\
				cB211.072.64 & 600000 & 750\textsf{x}512 &750\textsf{x}512&9000\textsf{x}32\\[-0.1cm]
                                & &  \,+\,deflation & &\\ 
				\botrule
			\end{tabular}
		}\\
	\caption{  Statistics used for the calculation of the disconnected contribution to the three-point function computed using two of the three  ensembles listed in Table~\ref{tab:params}. $C_{2pt}$ refers to the two-point function and $C^{f}_{\rm loop}$ refers to the fermion loop of flavor $f$. When a product is indicated hierarchical probing has been used having as first the number of stochastic vectors and as second the number of Hadamard vectors, i.e. $N_{\rm stoch}\textsf{x}N_{\rm Hadam}$. The notation ``+\,deflation'' means that the operator $C^{u+d}_{\rm loop}$ is deflated computing exactly 200 low-modes.}
	\label{tab:disconnected}
	\vspace*{-0.2cm}
\end{table}

\section{Analysis of correlators}
\label{sec:analysis}
The nucleon charges can be extracted by taking a ratio of $C^{f}_{\rm A,S,T}(t_s, t_{\rm ins})$ and $C_{\rm 2pt}(t_s)$ (c.f. Eqs.~\eqref{eq:thrp} and~\eqref{eq:twop}), 
\begin{equation}
R^{f}_{\rm A,S,T}(t_s,t_{\rm ins}) = \frac{C^{f}_{\rm A,S,T}(t_s,t_{\rm ins})}{C_{\rm 2pt}(t_s)} \mathrel{\mathop{\longrightarrow}^{\Delta E (t_s-t_{\rm ins})\gg 1}_{\Delta E t_{\rm ins} \gg 1}}g^f_{\rm A,S,T}
\label{eq:ratio}
\end{equation}
where  $\Delta E$ is the energy gap between the ground and first excited states.
This ratio becomes time independent for large values of $t_s$ and $t_{\rm ins}$ yielding a plateau, the value of which gives the desired  nucleon  charge, ${g^f_{\rm A,S,T}}$. In practice, $t_s$  cannot be chosen arbitrarily large because the statistical errors grow exponentially with $t_s$. Thus, we need to use the smallest  $t_s$ that ensures convergence to the nucleon state. In this work, we use several values of $t_s$ and   increase the statistics as we increase $t_s$   to keep the statistical error approximately constant, which is essential to reliably  assess excited states~\cite{Alexandrou:2017hac,vonHippel:2016wid}. In Table~\ref{tab:statistics} we give the values of $t_s$ used for the connected contribution and the associated statistics. A careful analysis is then performed, employing different methods to study ground state convergence. 

When fitting, we carry out correlated fits to the data, i.e.  we
compute the covariance matrix $v_{ij}$ between jackknife or bootstrap
samples and minimize
\begin{equation}
\chi_c^2 = \sum_{i,j}[y_i - f(\vec{b},x_i)] v^{-1}_{ij} [y_j - f(\vec{b}, x_j)],
\label{eq:chisq fit cova}
\end{equation}
where $y_i$ are the lattice data,
$f(\vec{b},x_i)$ is the fit function, which depends
on $x_i$ and $\vec{b}$. $x_i\in \{t_s,t_\textrm{ins}\}$ are the values of $t_\textrm{ins}$ and/or $t_s$ at which $y_i$ is evaluated and
$\vec{b}$ is a vector of the parameters being fitted for. 

\subsection{Plateau method}
\begin{figure}[h!]
	\centering
	\includegraphics[height=4.5cm]{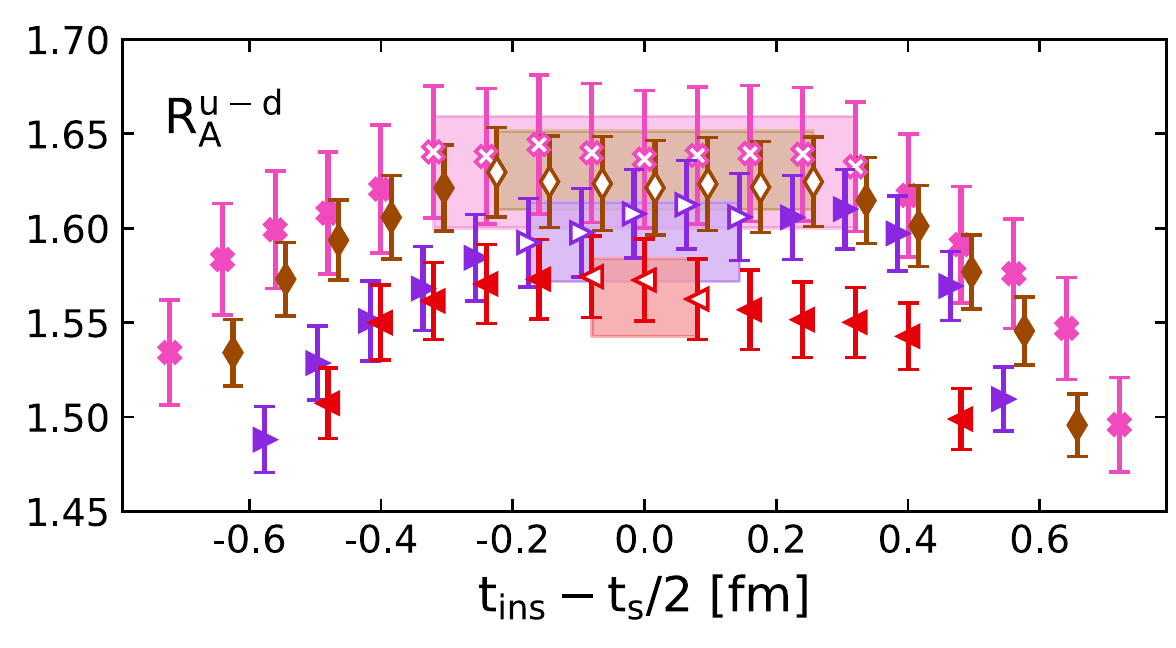}\\
	\vspace*{-0.3cm}
	\caption{\label{fig:plateau} We show the ratio of Eq.~\eqref{eq:ratio} as a function of $t_{\rm ins}$ for four $t_s$ values, namely $t_s/a = 14,16,18,$ and~$20$, where   the two-point function having the same statistics as the three point function. The ratio  yields, in the large Euclidean time limit, 
		the  isovector axial charge, $g_A^{u-d}$, for the cB211.072.64 ensemble. The bands and open symbols  mark the interval $t_\textrm{ins}\in[\tau_\textrm{plat}, t_s-\tau_\textrm{plat}]$ where the plateau fit is performed. 
	}
\vspace*{-0.2cm}
\end{figure}
We fit the ratio of Eq.~\eqref{eq:ratio} to a constant in an interval $t_\textrm{ins}\in[\tau_\textrm{plat}, t_s-\tau_\textrm{plat}]$. This assumes the ground state is the dominant contribution. We
choose $\tau_\textrm{plat}$ such that a constant fit describes well the data. The fits are performed independently for each $t_s$ using the same $\tau_\textrm{plat}$  and thus we fit only the data for which  $t_s$  satisfies $t_s>2\tau_\textrm{plat}$. We seek convergence of the fitted value  as $t_s$ increases. 
An example of this analysis, which we will refer to as the plateau method, is shown in Fig.~\ref{fig:plateau}, applied to the isovector axial charge, where we take $\tau_{\rm plat}/a=6$. As can be seen, as we increase $t_s$ the ratio increases, indicating a convergence only for the two largest time separations. However, the errors also increase and make it difficult to judge convergence.

\subsection{Two- and three-state fit}
In the two- and three-state fit approaches, we take into account the contributions of the first and second excited states, respectively, using  multiple  values of $t_s$ and fit them simultaneously.
The two-point correlator is described by the tower of states,
\begin{equation}\label{eq:corr_2pt}
C_{\rm 2pt}(t_s) = \sum_{i=0}^{\infty} c_i e^{-E_i t_s},
\end{equation}
where $E_0 = m_N$ is the nucleon mass and $E_i>m_N$ for $i>0$ are excited states with increasing energies. The amplitudes $c_i$ are positive numbers. In our two- and three-state fit analysis we fit the two-point functions using Eq.~\eqref{eq:corr_2pt} with, in the first case, $i=0$ and~$1$ and having four fit parameters and, in the latter case,  $i=0,~1,$~and $2$ and having six fit parameters.

The three-point function correlator is described by the tower of states,
\begin{align}\label{eq:corr_3pt}
C_{\rm 3pt}(t_s,t_{\rm ins}) = \sum_{i=0, j=0}^{\infty} A_{ij} e^{-E_i(t_s-t_\textrm{ins})}e^{-E_jt_\textrm{ins}}
\end{align}
where $A_{ij}$ are matrix elements and overlaps and $A_{ij} = A_{ji}$.
When we fit the three-point function with Eq.~\eqref{eq:corr_3pt}, we use the nucleon mass and the energies of the first and second excited states as  obtained from the fit of the two-point functions of Eq.~\eqref{eq:corr_2pt} and fit for the amplitudes $A_{ij}$.
The nucleon mass and energies are extracted via a jack-knife analysis from the two-point functions and the resampled values are subsequently used in the jack-knife analysis  of the three-point functions. Within such an approach all the nucleon matrix elements share the same set of energies and we restrict the search-space of the fit to the three-point function to the amplitudes $A_{ij}$ only.
This means that in the case of the two-state fit analysis and using Eq.~\eqref{eq:corr_3pt} with $i,j=0$ and~1, we have  three additional fit parameters to determine. For the three-state fit analysis we have $i,j=0,1,$ and~$2$ and six additional fit parameters.

\begin{figure}[h!]
	\centering
	\includegraphics[height=4.5cm]{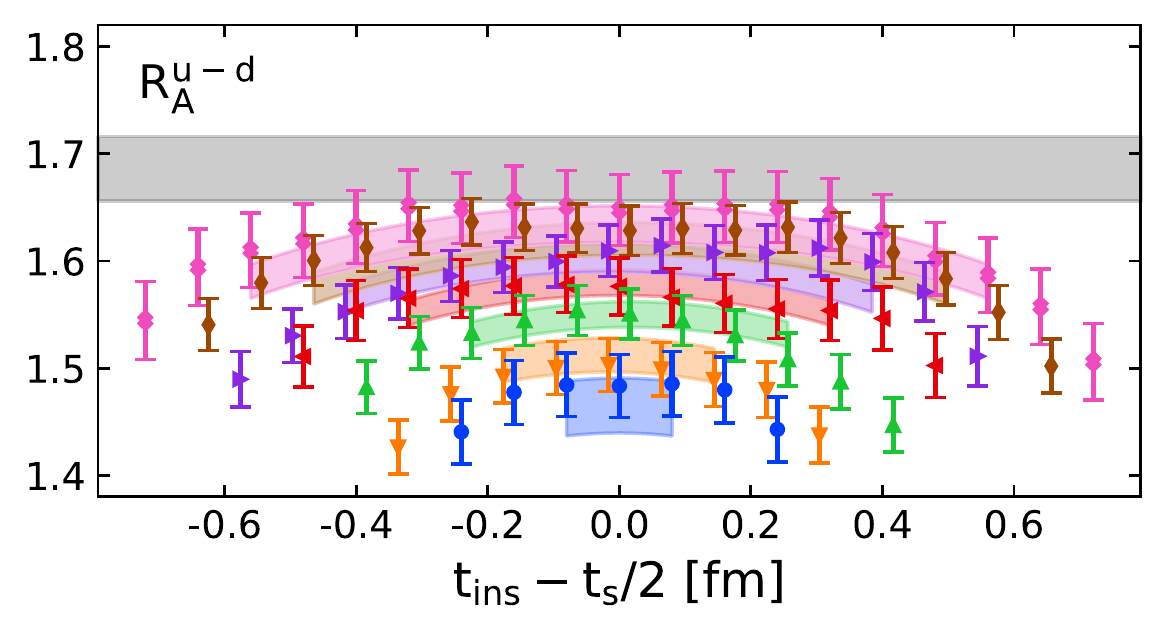}\\
	\vspace*{-0.3cm}
	\caption{\label{fig:two-state bands} The ratio of Eq.~(\ref{eq:ratio}) as a function of $t_{\rm ins}$ for various values of $t_s$, where we divide with the two-point function with the maximum statistics since this is what is fitted to extract the energies of the ground and first excited states.   We show  the associated bands resulting from a two-state fit for the cB211.072.64 ensemble. The grey band shows the extracted value of the bare $g_A^{u-d}$ from the fit. }
	\vspace*{-0.2cm}
\end{figure}

We demonstrate the application of the two- and three-state fit approaches in the case of the axial charge for the cB211.072.64 ensemble. 
We show in Fig.~\ref{fig:two-state bands} the ratio for each value of $t_s$ with the predicted curve from two-state fit obtained by fitting two- and three-point functions as described above. The extracted value of $g_A^{u-d} = A_{00}/c_0$ is shown with a gray band and the predicted curve for each $t_s$ is shown with a band of the same color as the points. 
Since we fit two-point functions with the largest statistic available, i.e. obtained by averaging forward, backward, neutron and proton nucleon correlators over 264 source positions per configuration, the ratios shown in Fig.~\ref{fig:two-state bands} are constructed by dividing  the three-point functions by the two-point function with the maximum statistics. This differs from the ratio used in the analysis of the plateau averages where we divide the three-point function by the  two-point function having the same statistics as the three point function to take advantage of the correlations between them.

\begin{figure}[h!]
	\centering
	\hspace*{-0.1cm}\includegraphics[height=4.5cm]{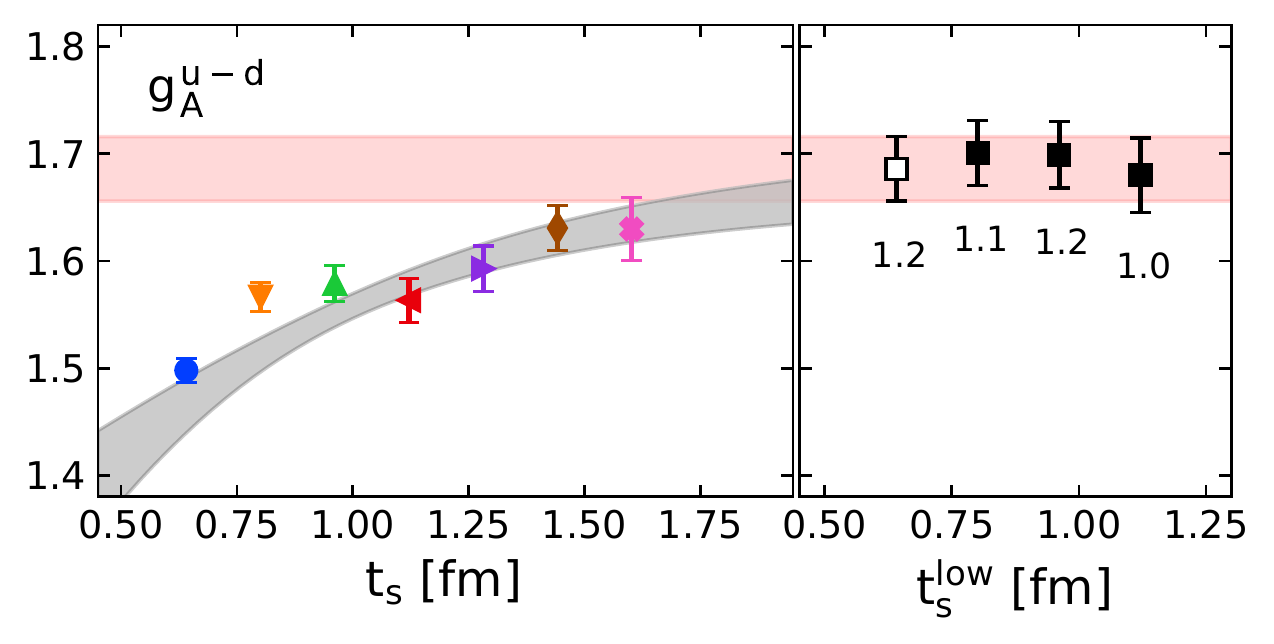}\\
	\vspace*{-0.3cm}
	\caption{\label{fig:two-state}  The  left panel shows the values extracted from either taking the mid-point of the ratio or from fitting the plateau to a constant as illustrated in Fig.~\ref{fig:plateau} for each $t_s$ value. We use the same symbols  at each $t_s$ as the symbols used for the ratio in Fig.~\ref{fig:two-state bands}. The gray band  shows the predicted curve as a function of $t_s$ if we use the parameters extracted from the selected two-state fit and taking $t_{\rm ins}=t_s/2$. The right panel shows the value of the bare $g_A^{u-d}$ extracted from the two-state fit (black squares) as a function of the lowest value  of $t_s$ included in the two-state fit, $t^{\rm low}_s$. The numbers below each point are the values of $\chi^2_c/{\rm d.o.f.}$ from the fit to the three-point function. The $\chi^2_c/{\rm d.o.f.}$ from the fit to the two-point functions using the two-state fit is the same for all values of $t_s$  and equal to $1.0$. The selected value for the two-state fit is marked with an open symbol and the red band shows the associated error across both panels.}
	\vspace*{-0.2cm}
\end{figure}

In Fig.~\ref{fig:two-state} we show the resulting curve using the parameters determined from the two-state fit as a function of $t_s$ fixing $t_{\rm ins}=t_s/2$ in 
Eq.~\eqref{eq:ratio}. We also show for each $t_s$ the extracted plateau value extracted, following the procedure explained in connection to  Fig.~\ref{fig:plateau}. For $t_s<2\tau_{\rm plat}$ we show the mid point of the ratio since no fit is performed.
As can be seen, the two-state fit predicts well the
behavior of the ratio and demonstrates that the asymptotic value is reached for values of $t_s$ larger than 2~fm, which is in agreement with the chiral perturbation analysis of Ref.~\cite{Bar:2018xyi}.
Therefore, taking the plateau value for the largest $t_s$ considered, which, within errors, seems to have converged,  underestimates $g_A^{u-d}$. In particular, notice that since data are correlated, one might mistake convergence in the window of $t_s \in 0.75-1.25$, which will lead to a small value of $g_A^{u-d}$. This is why it is important to have precise data for larger time separations.
In the right panel of Fig.~\ref{fig:two-state} we show the extracted value of $g_A^{u-d}$ as a function of $t_s^{\rm low}$, i.e. the smallest $t_s$ included in the  two-state fit. As can be seen, the values remain  consistent as $t_s^{\rm low}$  increases. 
We also give for each point the $\chi^2_c/{\rm d.o.f.}$ as given in Eq.\eqref{eq:chisq fit cova} that is close to unity for all the cases. 
We note that the $\chi^2_c$  is computed from the fit of the three-point function correlator and the quality of the fit can be better understood by looking at the curves depicted in Fig.~\ref{fig:two-state bands} than how well the gray band describes the plateau averages in Fig.~\ref{fig:two-state}, which are not fitted. 
Since the values show convergence as $t_s^{\rm low}$ increases  we select the value obtained by using $t_s^{\rm low} = 0.64$~fm. As we will see below,  this will be also in agreement with the three-state fit and the summation method and thus will fulfill our criterion for selecting the two-state fit value that also agrees with the value extracted from the summation method.

\begin{figure}[h!]
	\centering
	\hspace*{-0.1cm}\includegraphics[height=4.5cm]{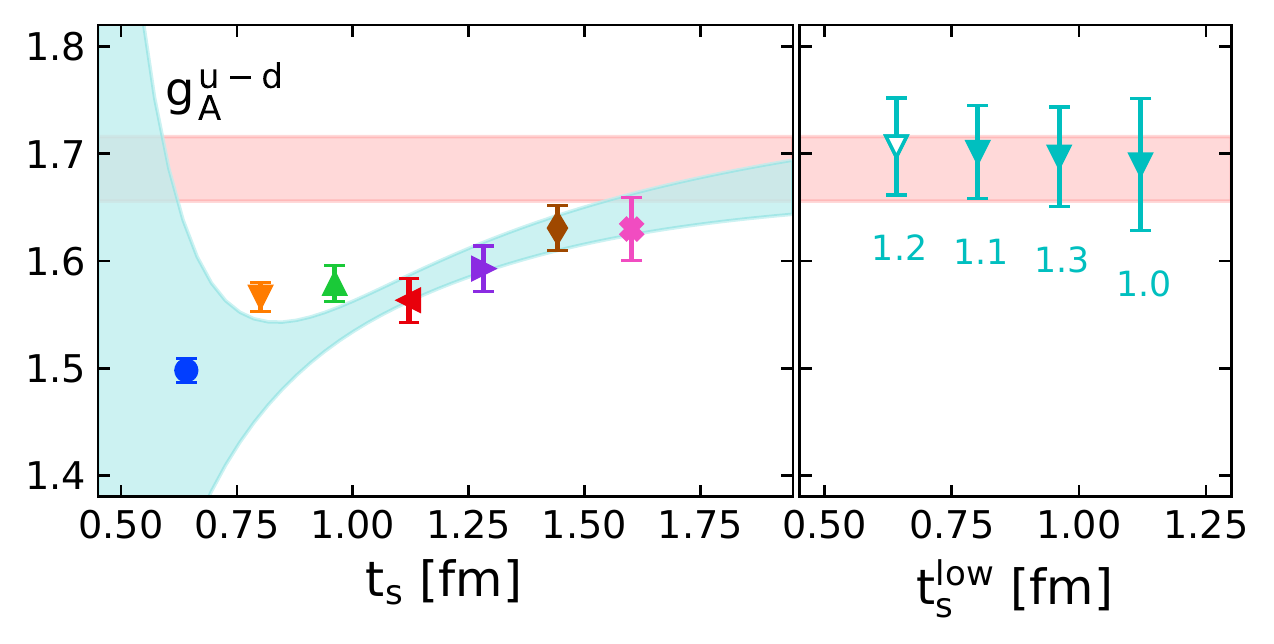}\\
	\vspace*{-0.3cm}
	\caption{\label{fig:three-state} The same as Fig.~\ref{fig:two-state} but using a three-state fit to the ratios of Eq.~(\ref{eq:ratio}) for the cB211.072.64 ensemble. The $\chi^2_c/{\rm d.o.f.}$ from the fit of the two-point functions using the three-state fit is the same for all values of $t_s$  and equal to $1.2$. The cyan band on the left  shows the predicted curve if we use the
		parameters extracted from the selected three-state fit marked with an open symbol. The red band shows for comparison the selected value from the two-state fit.}
	\vspace*{-0.2cm}
\end{figure}

A similar analysis is performed for the three-state fit approach, i.e. when two excited states are considered taking $i,j=0,1,$ and~$2$ in Eqs.~\eqref{eq:corr_2pt} and ~\eqref{eq:corr_3pt}. 
The results are given in Fig.~\ref{fig:three-state} where we use the same convention as for the two state-fit and for comparison we show with a red band the errors related to the selected two-state fit value.

\subsection{Summation method}
\begin{figure}[h!]
	\centering
	\includegraphics[height=4.5cm]{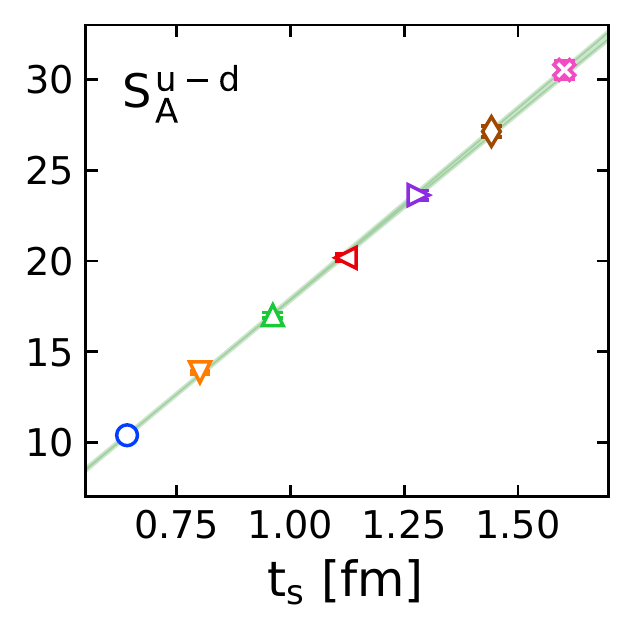}\includegraphics[height=4.5cm]{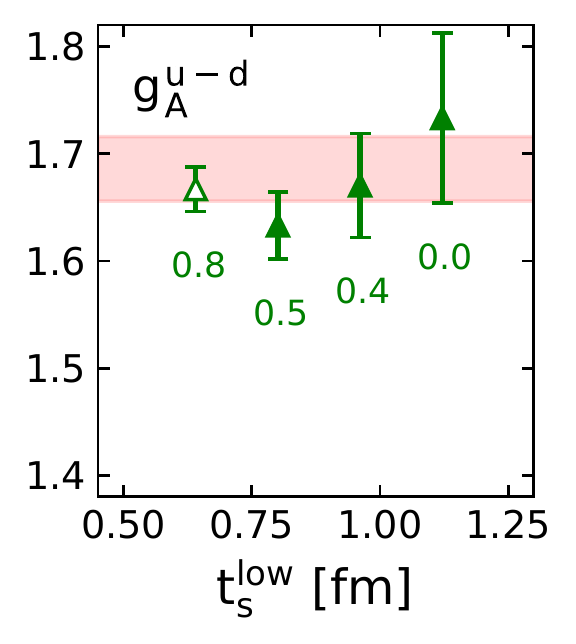}\\
	\vspace*{-0.3cm}
	\caption{\label{fig:summation} The left panel shows the summed ratio of Eq.~\eqref{eq:Summ} as a function of $t_s$ for the cB211.072.64 ensemble. The green band shows the resulting  curve using the open symbols  in the fit. The right panel shows the extracted value from the linear fit to the summed ratio (green up-triangles) as a function of the lowest value of $t_s$ included in the fit, $t^{\rm low}_s$. The numbers below each point are the values of $\chi^2_c/{\rm d.o.f.}$ from the fit. The open symbol shows the selected value  of the slope yielding $g_A^{u-d}$. The red band shows for comparison the result for the selected $t_s^{\rm low}$ from the two-state fit.}
	\vspace*{-0.4cm}
\end{figure}

We sum over the insertion time $t_{\rm ins}$ in the ratio in Eq.~(\ref{eq:ratio}) assuming only the lowest state dominates to obtain  a linear dependence on $t_s$, given by
\begin{equation}
S^{f}_{\rm A,S,T}(t_s) = \sum_{t_{\rm ins}=a}^{t_s-a} R(t_s,t_{\rm ins})\mathrel{\mathop{\longrightarrow}^{\Delta E t_s\gg 1}_{}}c + g^{f}_{\rm A,S,T}\,t_s\,, 
\label{eq:Summ}
\end{equation}
where in the sum we omit the source and sink time slices. The slope extracted from a linear fit to Eq.~(\ref{eq:Summ}) gives the nucleon charge in the limit of large $t_s$.
By increasing the lowest value of $t_s$ used in the fit we look for convergence in the extracted slope.  The
advantage of the summation method is that, despite the fact that it
still assumes a single state dominance, the excited states are
suppressed exponentially with respect to $t_s$ instead of $t_s-t_{\rm
	ins}$ that enters in the plateau method~\cite{Capitani:2012gj}. On the other hand, the
errors tend to be larger since we have two parameters to fit.
As an example we show in Fig.~\ref{fig:summation} the results after applying the summation approach for the case of  the isovector axial charge using the cB211.072.64 ensemble. 
We  select as our final value the one extracted from the two-state fit when excited states are detected, since the two-state fit models the data better than either the plateau or  the summation method in these cases. Even though the statistical error on the extracted value of the charges is larger than that extracted using the summation method,
  we prefer to be conservative so that we do not underestimate errors.

\subsection{Analysis of the cB211.072.64 ensemble}

\begin{figure}
	\centering
	\includegraphics[height=17cm]{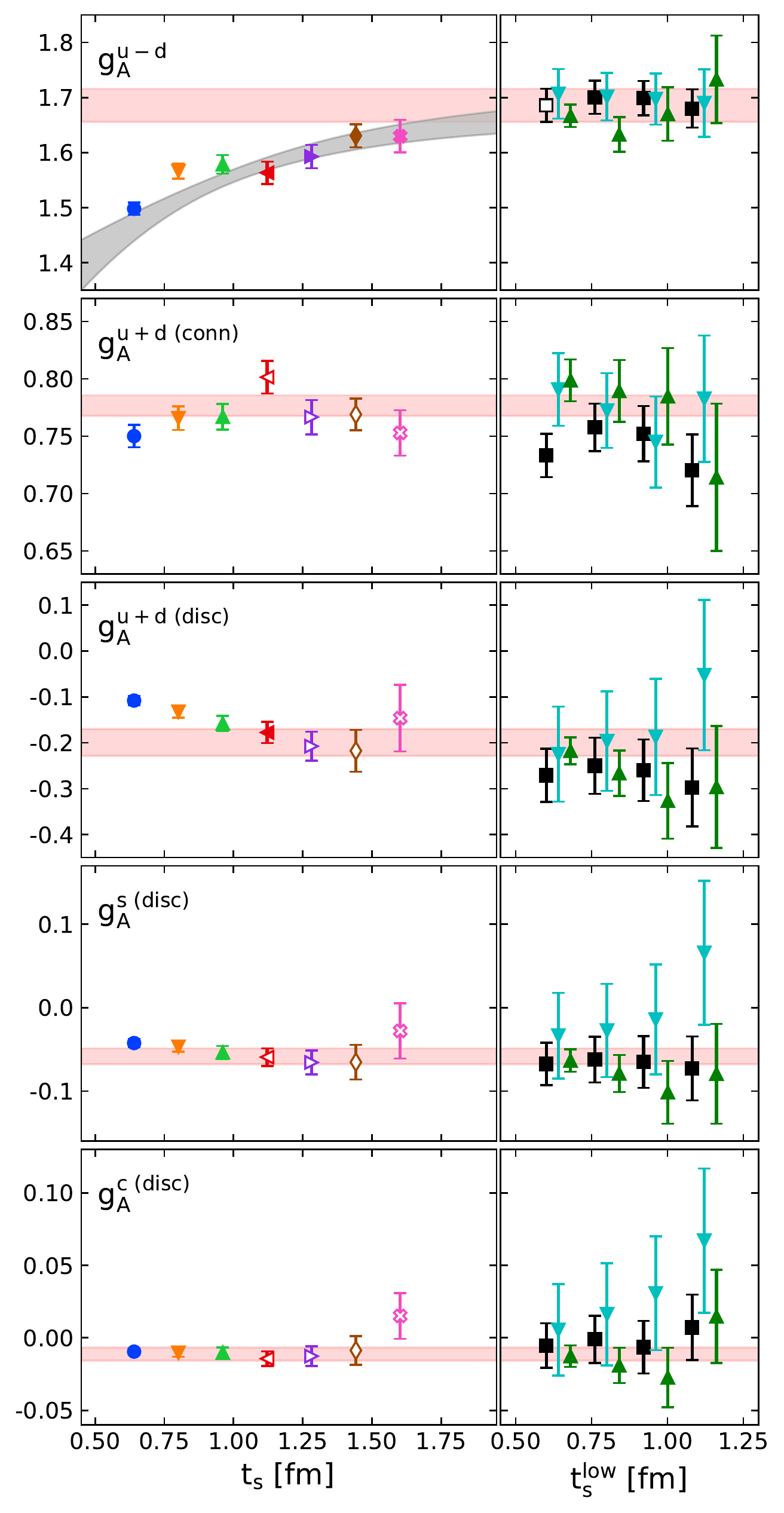}\\
	\vspace*{-0.3cm}
	\caption{\label{fig:cB211.64_gA} We show results for the  connected and disconnected contributions to the axial charges $g_A$ for the cB211.072.64 ensemble. The  left panels show the values extracted from either  fitting the plateau to a constant as illustrated in Fig.~\ref{fig:plateau} for each $t_s$ value when $t_s>2\tau_{\rm plat}$ or taking the midpoint for other values of $t_s$. We use the same colors and symbols  at each $t_s$ as those used for the ratio in Fig.~\ref{fig:plateau}. The  right panel shows the extracted value of $g=A_{00}/c_0$ from the two-state fit (black squares), three-state fit (cyan downwards-pointing triangles) and summation (green upwards-pointing triangles) as a function of the lowest value  of $t_s$ included in the two- and three-state fits, $t^{\rm low}_s$. 
	The red band across both panels shows the associated error to the selected final value. When a two-state fit value is selected, we mark on the right panel the selected value with open symbol and we show with the grey band the predicted curve for the ratio in Eq.~\eqref{eq:ratio} fixing $t_{\rm ins} = t_s/2$. When a plateau average is selected as final value, we mark on the left panel with open symbols the selected interval in $t_s$ where the plateau average is performed.}
\vspace*{-0.4cm}
\end{figure}

In Fig.~\ref{fig:cB211.64_gA}  we show a comparison among the three aforementioned
analysis methods for the axial charges using the cB211.072.64 ensemble. As already observed, the  value extracted for the axial charge $g_A^{u-d}$ from the two-state fit shows very mild dependence on the $t_s^{\rm low}$ used in the fits. In addition, the value extracted from the two-state fit is confirmed by the three-state fits as well as by the summation method. Therefore, we take as our final result the value extracted from the two-state fit  for which $\chi^2/{\rm d.o.f.} \sim 1$ and  there is agreement with  the summation method. The connected and the disconnected parts of the isoscalar charge $g_A^{u+d}$ as well as the strange and charm disconnected contributions do not suffer from
large excited states contamination, as can be seen in Fig.~\ref{fig:cB211.64_gA} by the fact that the plateau results at the largest three or four values of $t_s$ are roughly constant. 
Furthermore, the plateau values  converge to a  value that
is in agreement with the values extracted using two- and three-state fits and the summation methods and we thus take the weighted average.  We note that
we follow this procedure for all  cases where such a behavior is observed, i.e  we take the correlated average of the plateau values for the range of $t_s$ for which these have converged.

\begin{figure}
	\includegraphics[height=17cm]{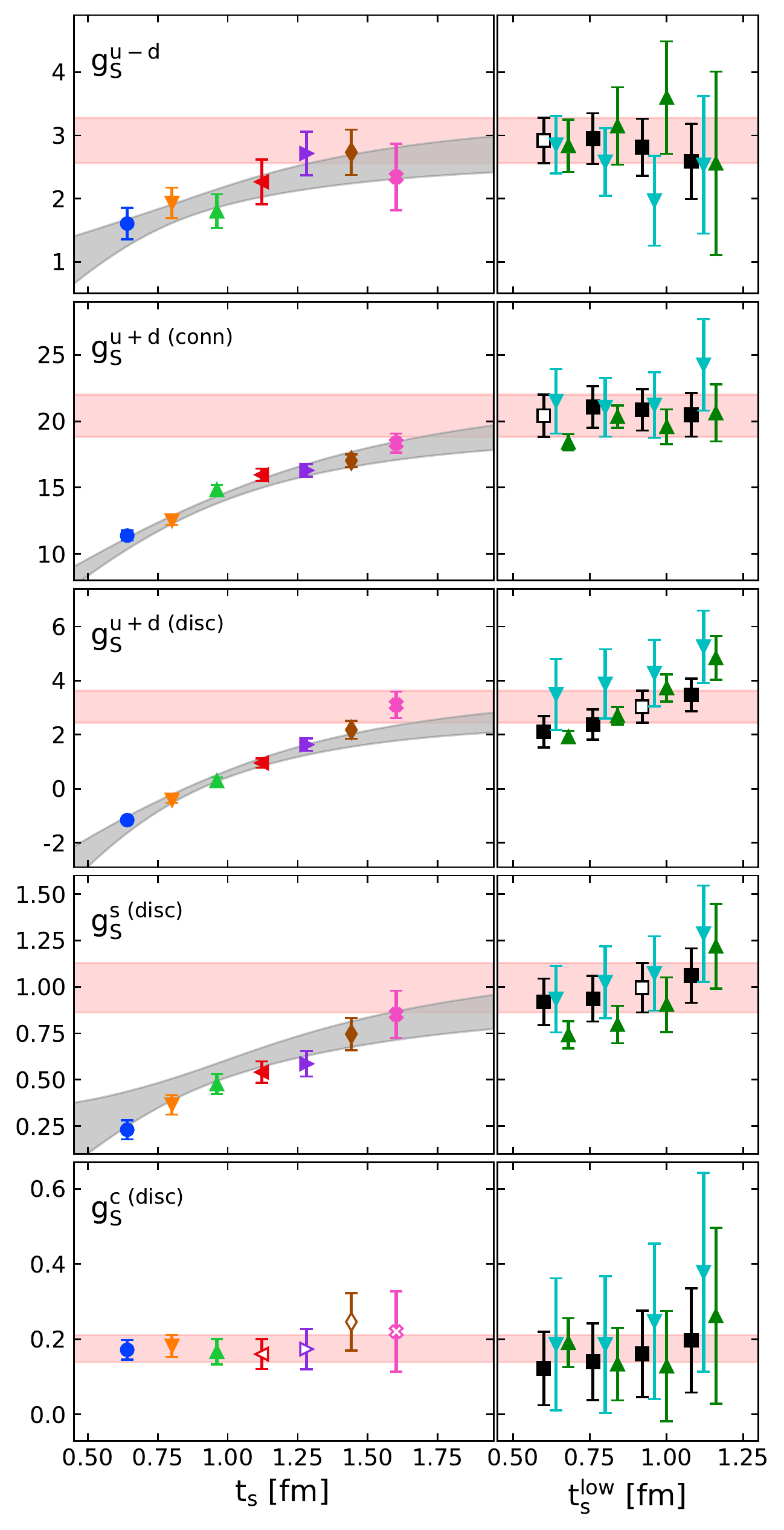}\\
	\vspace*{-0.3cm}
	\caption{\label{fig:cB211.64_gS} We show results for the connected and disconnected contributions to the scalar charges $g_S$ for the cB211.072.64 ensemble. The notation is the same as that in Fig.~\ref{fig:cB211.64_gA}.}
	\vspace*{-0.4cm}
\end{figure}

We perform the same analysis for the scalar charge and we show the results in Fig.~\ref{fig:cB211.64_gS}. Contrary to the axial charge, all the contributions to the scalar charges, except for the charm disconnected component, suffer from
large excited states contamination. As can be seen, the plateau values  show a similar behavior as $g_A^{u-d}$ showing convergence of the two-state fit at $t_s$ larger than 2~fm. We therefore choose as our final value  the one extracted from the two-state fit when it agrees with the summation method and does not show dependence on $t_s^{\rm low}$. We thus use $t_s^{\rm low}=8a=0.64$~fm in the fit of the isovector $g_S^{u-d}$, connected isoscalar $g_S^{u+d\ {\rm(conn)}}$ and the strange disconnected scalar charge $g_S^{s}$.  For the disconnected isoscalar scalar charge $g_S^{u+d\ {\rm(disc)}}$ and $g_S^{s{\rm(disc)}}$, we use a larger value $t_s^{\rm low}=0.96$~fm to better account for the upward trend still present in the data. We note that the results on the scalar charges have errors about ten times larger as compared to the axial charges.

\begin{figure}
	\includegraphics[height=17cm]{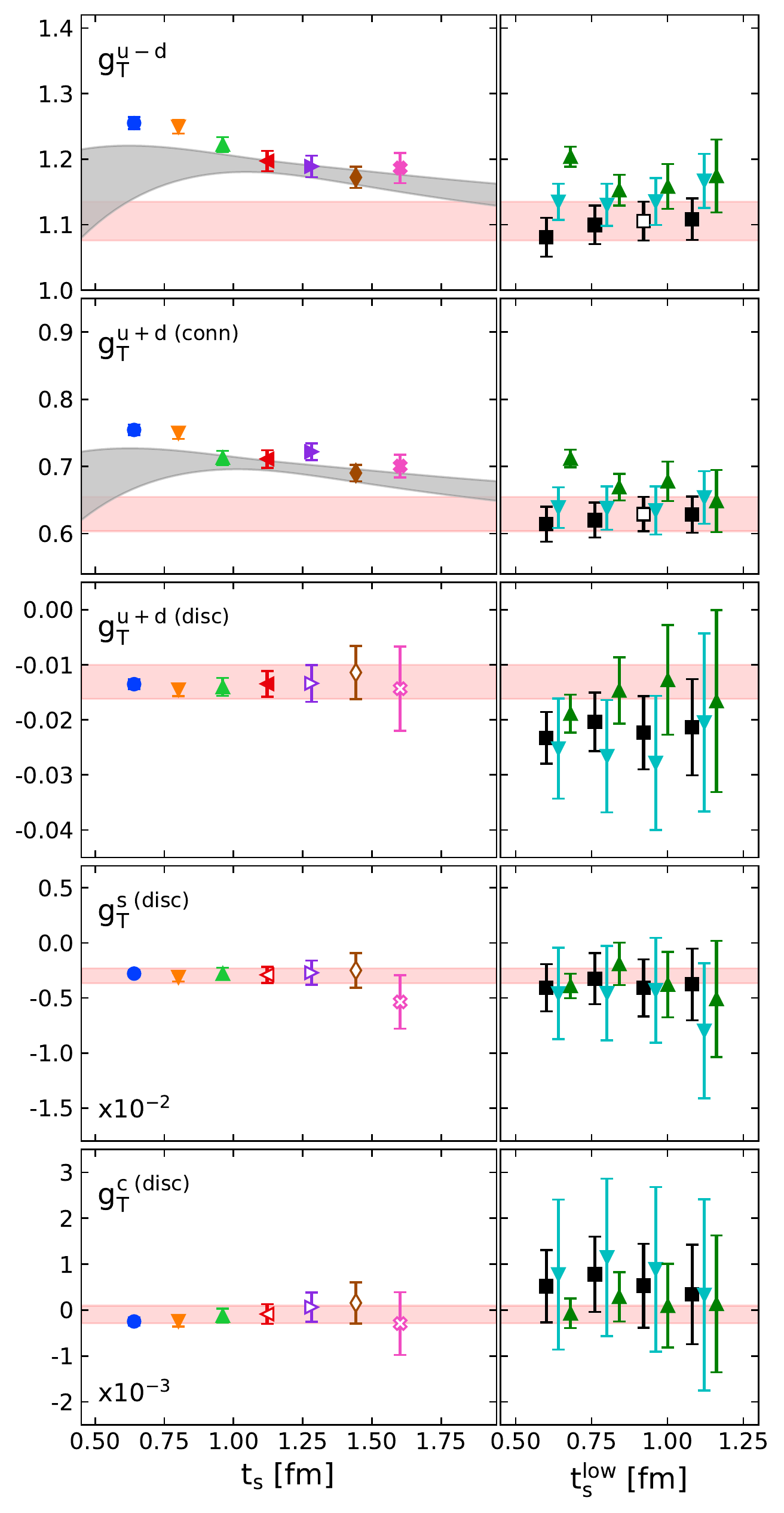}\\
	\vspace*{-0.3cm}
	\caption{\label{fig:cB211.64_gT} We show results for the connected and disconnected contributions to the tensor charges $g_T$ for the cB211.072.64 ensemble. The notation is the same as that in Fig.~\ref{fig:cB211.64_gA}.}
	\vspace*{-0.4cm}
\end{figure}

We perform the same analysis for the tensor charges and show the resulting values in Fig.~\ref{fig:cB211.64_gT}.  As can be seen, for the isovector $g_T^{u-d}$ and connected isoscalar tensor charge $g_T^{u+d\ {\rm(conn)}}$, there is no agreement between the  two-state and summation values for smaller $t_s^{\rm low}$ values. The value from the two-state fit
becomes consistent with the one extracted from the summation when $t_s^{\rm low}=12a=0.96$~fm and this is what we select as final value. This demonstrates that the larger values of $t_s$ are crucial for properly probing ground state dominance.

The results for the connected isoscalar charge $g_T^{u+d\ {\rm(conn)}}$ are shown in Fig.~\ref{fig:cB211.64_gT}. As can be seen, the plateau values computed by fitting to a constant using the data at $t_s=0.96,1.12$ and $1.28$~fm are compatible and one may think that the values have converged. However, both the results from the two-state and three-state fits are lower. Furthermore,  there is a tension between the two-state fit and the summation method for  $t_s^{\rm low}=0.64$~fm and only when  $t_s^{\rm low}=1.28$~fm that they become consistent. This would correspond to about 2.6~fm in the plateau method.  This again demonstrates the importance of having precise data for large $t_s$ values.

 Similarly to the axial charge, the disconnected contributions do not show 
large effects from excited states. We thus take the average of the plateau values for the range of $t_s$ for which these have converged as the final value.

\begin{table}
	{
		\centering
		\small
		\renewcommand{\arraystretch}{1.2}
		\renewcommand{\tabcolsep}{1pt}
		\begin{tabular}{l|ccccc}
			\toprule
			         & ${u\texttt{-}d}$ & ${u\texttt{+}d}$ (conn) & ${u\texttt{+}d}$ (disc) &     $s$      &     $c$      \\ \hline
			$g_A$    &    1.686(30)     &       0.7766(90)        &       -0.199(29)        & -0.0583(96)  & -0.0112(45)  \\
			$g_S$    &     3.04(59)     &        20.4(1.6)        &        3.04(59)         &   1.00(13)   &  0.175(36)   \\
			$g_T$    &    1.106(32)     &        0.629(27)        &       -0.0131(31)       & -0.00299(68) & -0.00010(19) \\
			\botrule 
		\end{tabular}
	}
	\caption{The nucleon axial, scalar and tensor charges extracted using the  cB211.072.64 ensemble.}
	\label{tab:charges_bare}
\end{table} 
We summarize in Table~\ref{tab:charges_bare} the bare values for the isovector, connected and disconnected isoscalar, strange and charm  axial, scalar and tensor charges.
These results clearly demonstrate that  disconnected contributions  cannot be neglected at the physical point. They are enhanced in comparison to the values obtained at heavier pion masses where for example the disconnected part of  $g_A^{u+d}$ using a $N_f=2+1+1$   ensemble simulated at a pion mass of $m_\pi=370$ MeV was  -0.07(1)~\cite{Abdel-Rehim:2013wlz} as  compared to -0.199(29) for the cB211.072.64 ensemble.

\subsection{Analysis of cA2.09.48}

\begin{figure}
	\includegraphics[height=17cm]{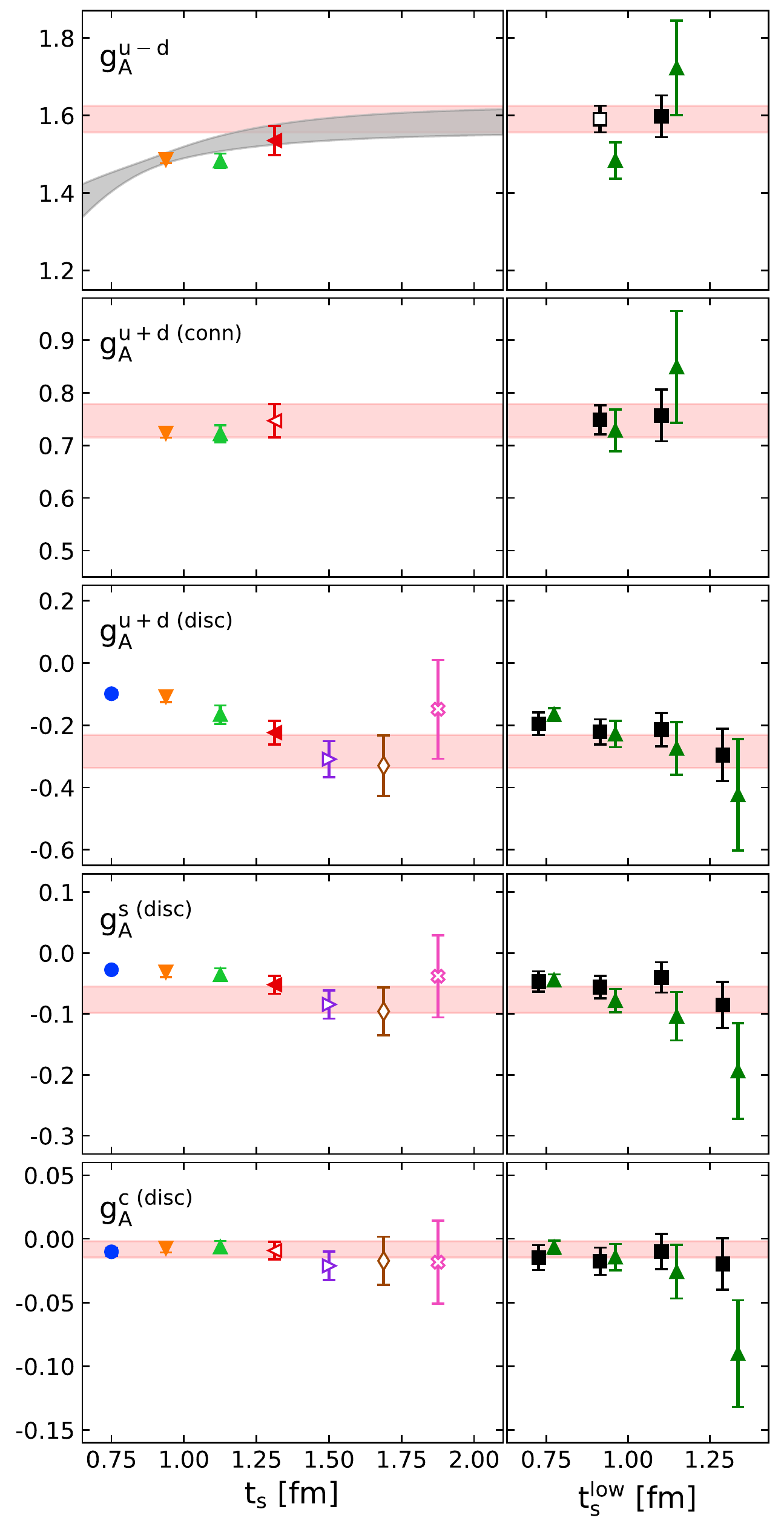}\\
	\vspace*{-0.3cm}
	\caption{\label{fig:cA2.48_gA} We show results for the connected and disconnected contributions to the axial charges $g_A$ for the cA2.09.48 ensemble. The notation is the same as that in Fig.~\ref{fig:cB211.64_gA}.}
	\vspace*{-0.4cm}
\end{figure}

\begin{figure}
	\includegraphics[height=17cm]{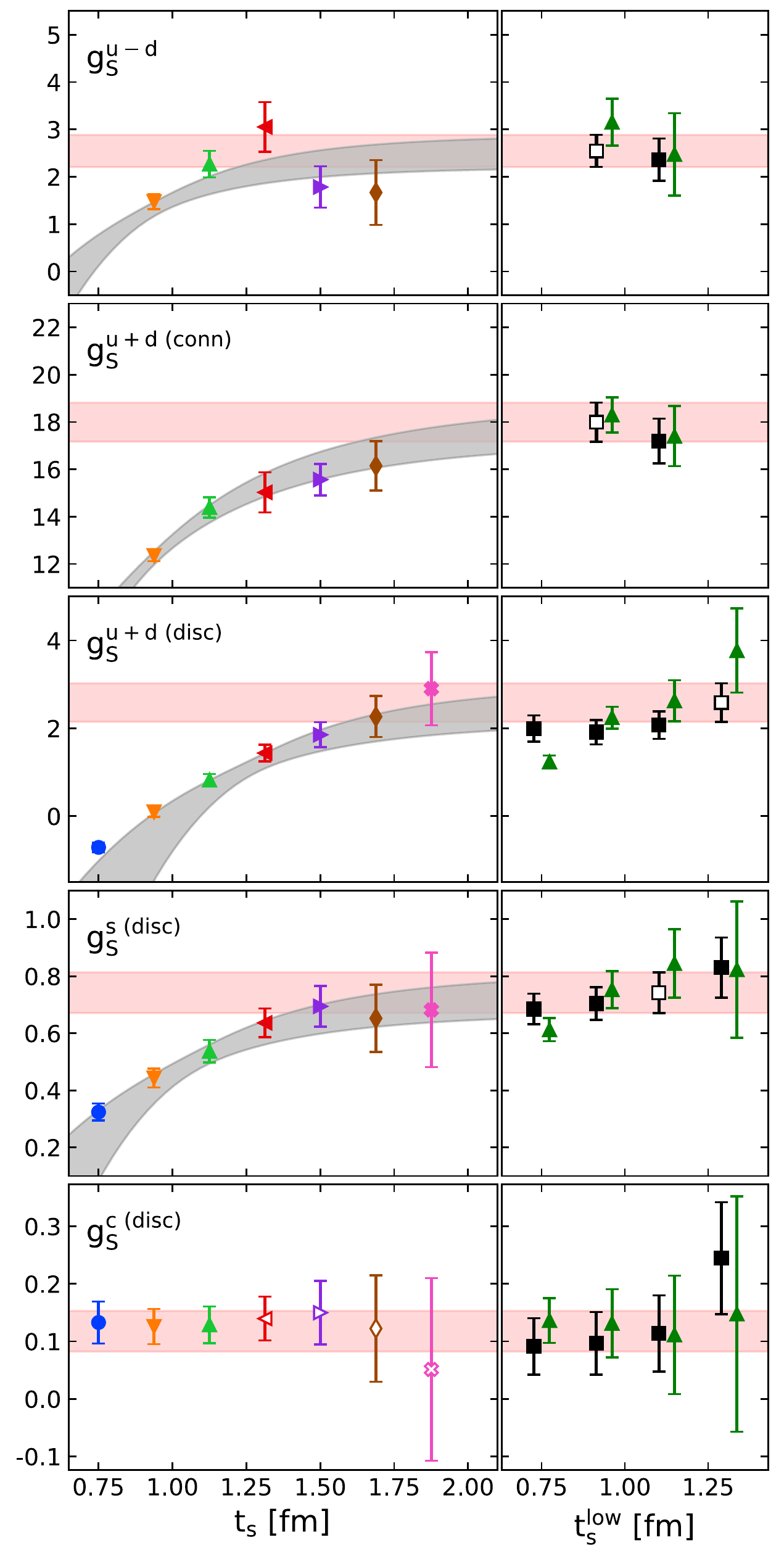}\\
	\vspace*{-0.3cm}
	\caption{\label{fig:cA2.48_gS} We show results for the connected and disconnected contributions to the scalar charges $g_S$ for the cA2.09.48 ensemble. The notation is the same as that in Fig.~\ref{fig:cB211.64_gA}.}
	\vspace*{-0.4cm}
\end{figure}

\begin{figure}
	\includegraphics[height=17cm]{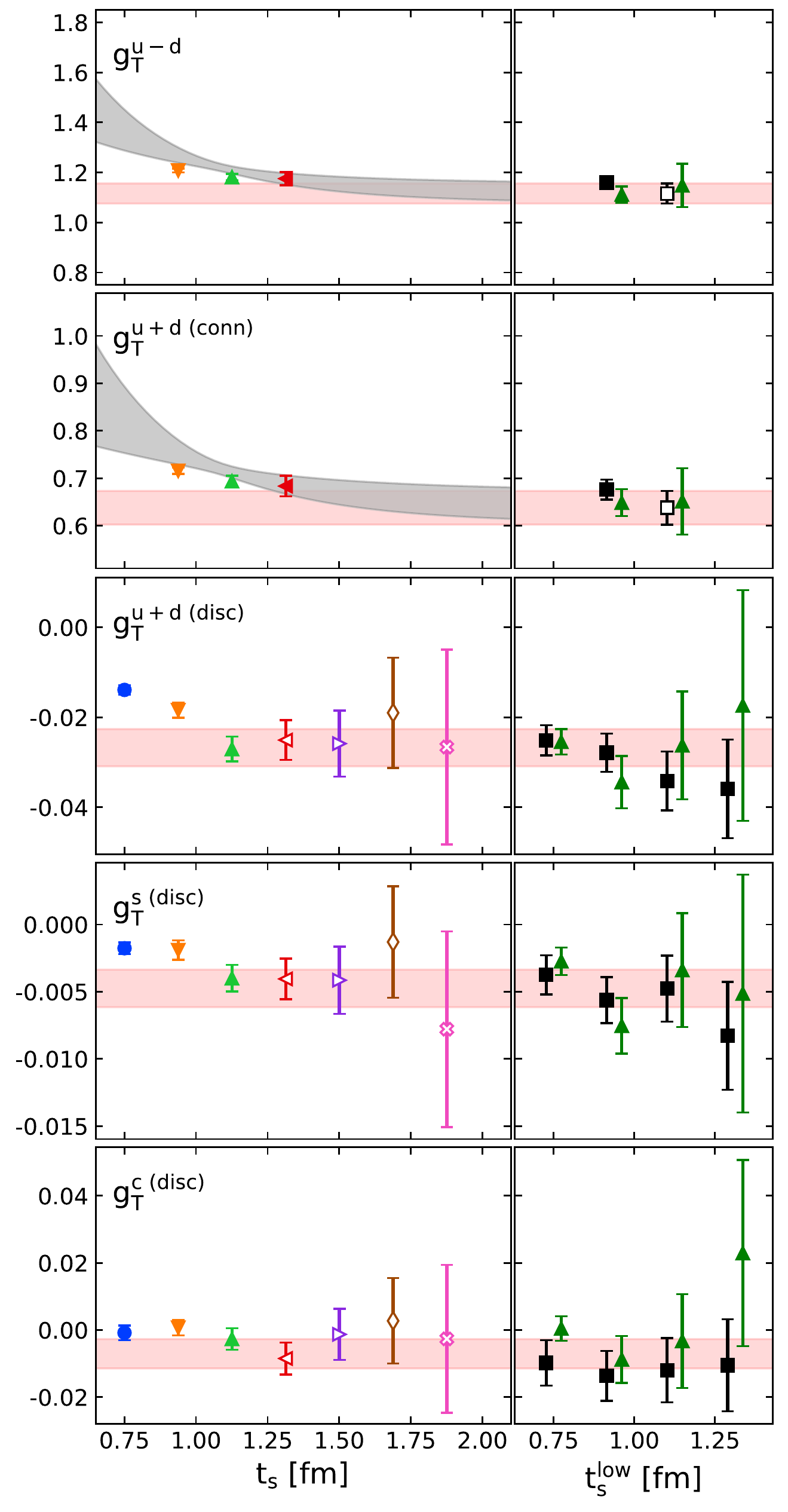}\\
	\vspace*{-0.3cm}
	\caption{\label{fig:cA2.48_gT} We show results for the connected and disconnected contributions to the tensor charges $g_T$ for the cA2.09.48 ensemble. The notation is the same as that in Fig.~\ref{fig:cB211.64_gA}.}
	\vspace*{-0.4cm}
\end{figure}

 We repeat the same analysis described for the cB211.072.64 ensemble for the cA2.09.48 ensemble. Since we now use 
 correlated fits the values presented in Refs.~\cite{Alexandrou:2017hac,Alexandrou:2017qyt} are modified but remain within their statistical errors. For the connected components on this ensemble we have only three values of $t_s$ for the matrix element of the axial and tensor currents determining  $g_A$ and $g_T$ and five for the scalar current determining $g_S$. The three smaller values of $t_s$ have constant statistics, as shown in Table~\ref{tab:statistics}, and thus their statistical errors increase significantly with increasing $t_s$. This means that the quality of the fits are not as good as for the cB211.072.64 ensemble. In particular, a three-state fit analysis cannot be performed  due to the low statistics and the small number of $t_s$. On the other hand the disconnected contributions are available for a larger number of $t_s$ and we thus show a more complete analysis.
 The results of the  analysis are summarized  in Fig.~\ref{fig:cA2.48_gA} for the axial charges, in Fig.~\ref{fig:cA2.48_gS} for the scalar charges and in Fig.~\ref{fig:cA2.48_gT} for the tensor charges. For the connected matrix elements of the axial and scalar current  the two-state fit with $t_s^{\rm low}=0.94$~fm agrees with the summation value as well as with the value extracted when using $t_s^{\rm low}=1.13$~fm and thus we take it as final value. On the other hand the tensor charge shows a more severe contamination of excited states and, similarly to the cB211.072.64 ensemble, we take as final value the two-state fit result at larger separation, namely $t_s^{\rm low}=1.13$~fm.
 Disconnected contributions to axial and tensor charges show very mild excited state contamination and we take the plateaus average as final value. On the hand for the scalar charge excited states are more severe and we take the two-state fit result at $t_s^{\rm low}=1.31$~fm for the disconnected isoscalar $g_S^{u+d}$ and $t_s^{\rm low}=1.13$~fm for the strange disconnected $g_S^{s}$.
 We summarize in Table~\ref{tab:charges_bare_cA2.48} the bare values for the isovector, connected and disconnected isoscalar, strange and charm  axial, scalar and tensor charges for this ensemble.
 
\begin{table}
	{
		\centering
		\small
		\renewcommand{\arraystretch}{1.2}
		\renewcommand{\tabcolsep}{1pt}
		\begin{tabular}{l|ccccc}
			\toprule
			& ${u\texttt{-}d}$ & ${u\texttt{+}d}$ (conn) & ${u\texttt{+}d}$ (disc) &     $s$      &     $c$      \\ \hline
			$g_A$    &    1.590(35)     &       0.747(32)        &       -0.284(53)        & -0.077(22)  & -0.0082(64)  \\
			$g_S$    &     2.54(34)     &        17.99(82)        &        2.59(44)         &   0.742(71)   &  0.118(35)   \\
			$g_T$    &    1.116(40)     &        0.638(35)        &       -0.0268(42)      & -0.0048(14) & -0.0071(44) \\
			\botrule 
		\end{tabular}
	}
	
	\caption{The nucleon axial, scalar and tensor charges extracted using the  cA2.09.48 ensemble.}
	\label{tab:charges_bare_cA2.48}
	%\vspace*{-0.2cm}
\end{table} 
 
\subsection{Analysis of cA2.09.64}
\begin{figure}[h!]
	\includegraphics[height=20.3cm]{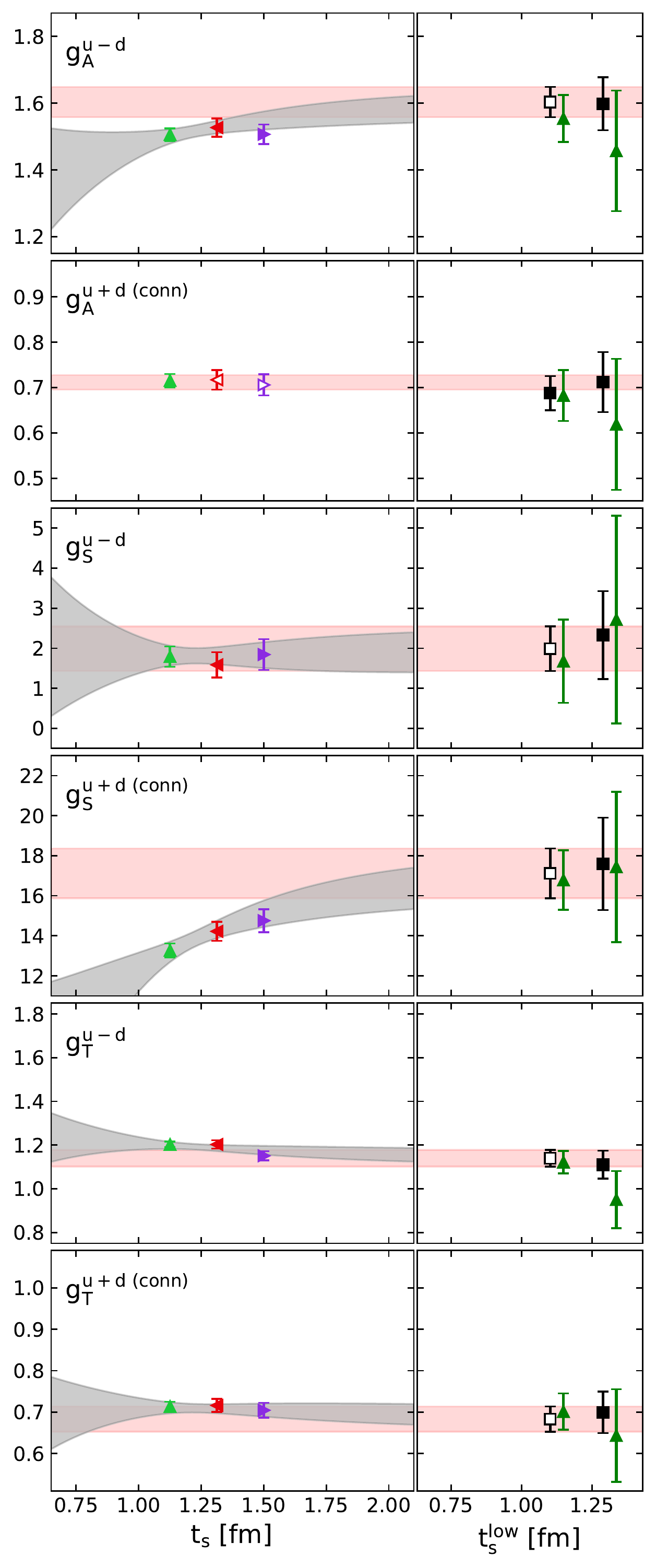}\\
	\vspace*{-0.3cm}
	\caption{\label{fig:cA2.64} We show results for the connected  contributions to the  axial $g_A$, scalar $g_S$ and tensor charges $g_T$  for the cA2.09.64 ensemble. The notation is the same as that in Fig.~\ref{fig:cB211.64_gA}.}
	\vspace*{-0.4cm}
\end{figure}
For the  cA2.09.64 ensemble we only have three values of $t_s$ and therefore the analysis of excited states is again not as accurate as for the cB211.072.64 ensemble. In addition we only have  connected contributions since the purpose of the analysis of the cA2.09.64 ensemble is to check for finite volume effects using the connected contributions which are much more precise and less expensive. Following the same analysis procedure we summarize the results on the isovector and connected isoscalar charges  in Fig~\ref{fig:cA2.64}. Given that the values for the two available $t_s^{\rm low}$ are consistent between them and with the values from the summation, we take as our selected values the ones extracted from the two state fits when using $t_s^{\rm low}=1.13$.  We remark that  while for scalar charge the plateau values show convergence we know from the more accurate analysis using the cB211.072.64 ensemble that this quantity has non-negligible contribution from excited states and thus we still fit it using a two-state fit. If we were to extract it using the weighted average plateau values we would obtain a value that is compatible with that from the two-state fit with however a smaller error. We therefore conservatively quote the value with the larger statistical error. 

We summarize in Table~\ref{tab:charges_bare_cA2.64} the bare values for the isovector and connected isoscalar  axial, scalar and tensor charges that can be directly compared to the values listed in Table~\ref{tab:charges_bare_cA2.48} for the cA2.09.48 ensemble, since these ensembles have the same renormalization constants. The results obtained using the cA2.09.48 and cA2.09.64 are compatible indicating that finite size effects are within  our statistical accuracy.

\begin{table}
	{
		\centering
		\small
		\renewcommand{\arraystretch}{1.2}
		\renewcommand{\tabcolsep}{1pt}
		\begin{tabular}{l|cc}
			\toprule
			& ${u\texttt{-}d}$ & ${u\texttt{+}d}$ (conn) \\ \hline
			$g_A$    &    1.603(45)     &       0.711(16)        \\
			$g_S$    &     1.99(56)     &        17.1(1.2)        \\
			$g_T$    &    1.139(38)     &         0.683(30)        \\
			\botrule 
		\end{tabular}
	}
	
	\caption{The nucleon axial, scalar and tensor charges extracted using the  cA2.09.64 ensemble.}
	\label{tab:charges_bare_cA2.64}
	\vspace*{-0.2cm}
\end{table} 
%\medskip
%\noindent
\section{Renormalization}
\label{sec:renormalization}
 Lattice QCD matrix elements  must be renormalized to extract physical quantities. We use  the ${\rm RI^\prime_{MOM}}$ scheme~\cite{Martinelli:1994ty} to compute non-perturbatively the renormalization functions using the momentum source method~\cite{Gockeler:1998ye}. We implement it as done in Ref.~\cite{Alexandrou:2012mt} and remove lattice spacing effects by subtracting  ${\cal	O}(g^2\,a^\infty)$ terms computed in perturbation theory~\cite{Constantinou:2009tr,Alexandrou:2015sea}.
We  distinguish between non-singlet and singlet renormalization functions, where for the latter we  compute, in addition to the connected, the disconnected contributions. The non-singlet and singlet  renormalization functions $Z_A$, $Z_P$ and $Z_T$ for the $N_f=2+1+1$ ensemble cB211.072.64 are computed using $N_f=4$ ensembles simulated at the same $\beta$ value and at five values of the pion mass so the chiral limit can be taken. The parameters for these ensembles are given in
Table~\ref{tab:Z_ensembles}. For the renormalization of the matrix elements using the  cA2.09.48 and cA2.09.64 ensembles we have analyzed $N_f=2$ ensembles as  
 extensively discussed in Ref.~\cite{Alexandrou:2015sea}. The scalar quantities are renormalized with the pseudo-scalar renormalization constant, $Z_P$, since they are computed using the pseudo-scalar current in the twisted-mass formulation.

\begin{table}[!h]
	\label{tab:Z_ensembles}  
	{
		\centering
		\small
		\renewcommand{\arraystretch}{1.2}
		\renewcommand{\tabcolsep}{1.5pt}
		\begin{tabular}{ccc}
			\hline
			\hline
			$\,\,\,$        $\,\,\,$              & $\beta=1.778$, $\,\,\,a=0.08$ fm  &          \\ \hline
			$a \mu$   & $a m_{\pi}$ & lattice size\\
			\hline
			$\,\,\,$  0.0060$\,\,\,$         & 0.14836   & $24^3 \times 48$ \\  
			$\,\,\,$  0.0075$\,\,\,$         & 0.17287   & $24^3 \times 48$ \\          
			$\,\,\,$  0.0088$\,\,\,$         & 0.18556   & $24^3 \times 48$ \\  
			$\,\,\,$  0.0100 $\,\,\,$        & 0.19635   & $24^3 \times 48$ \\ 
			$\,\,\,$  0.0115 $\,\,\,$        & 0.21028   & $24^3 \times 48$  \\ 
			\hline
			\hline
		\end{tabular}
	}
	\caption{Parameters for $N_f=4$ ensembles needed for the renormalization
	of the cB211.072.64 ensemble ($N_f=2+1+1$). $\mu$ is the twisted mass parameter.}
	\vspace*{-0.2cm}
\end{table}

\begin{figure}
	\includegraphics[width=1\linewidth]{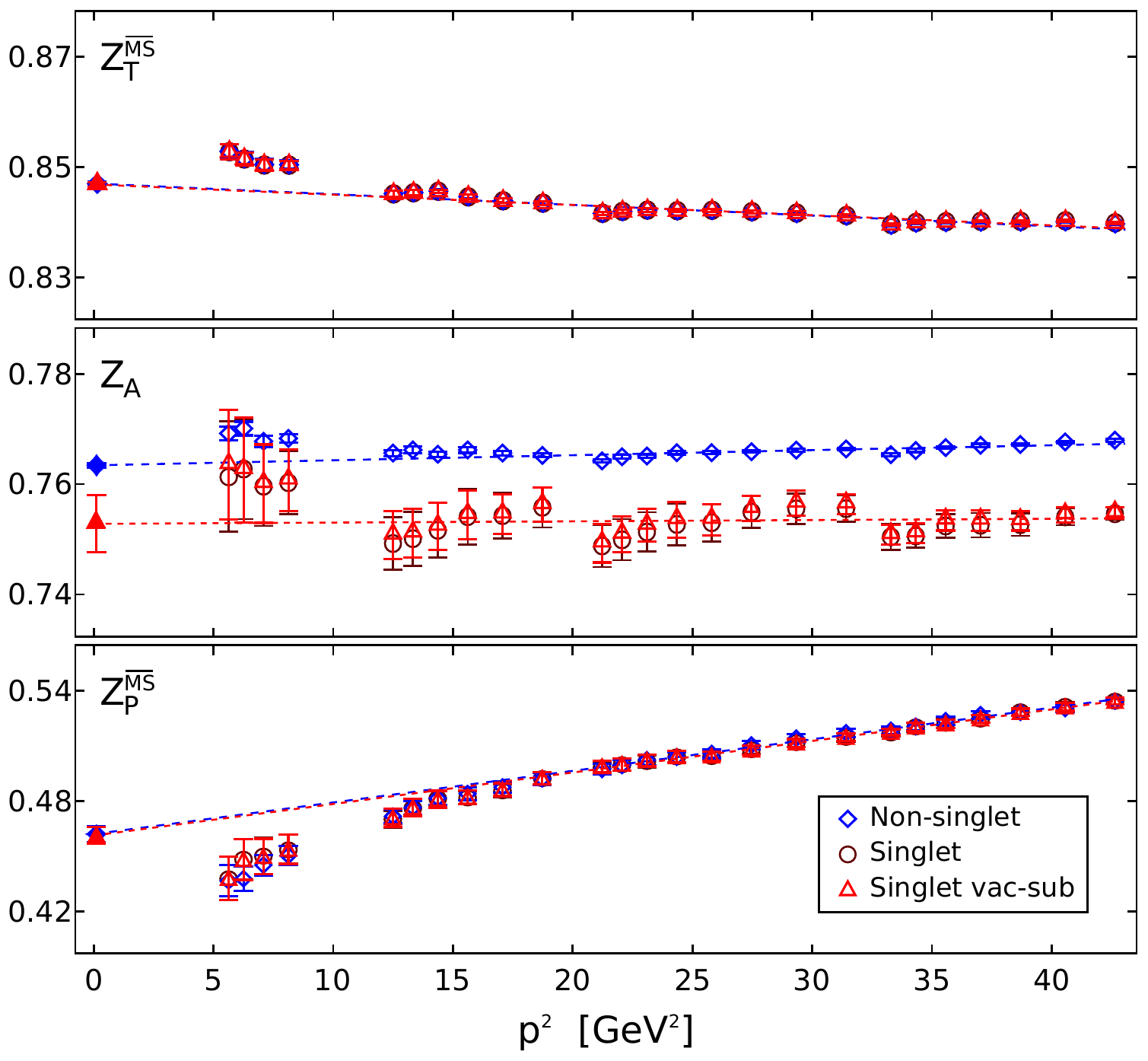}\\
	\vspace*{-0.3cm}
	\caption{\label{fig:Z} Non-singlet and singlet renormalization functions computed on a $N_f=4$ ensemble with the same $\beta$-parameter of the cB211.072.64 ensemble. }
	\vspace*{-0.2cm}
\end{figure}

We show in Fig.~\ref{fig:Z} the determination of the non-singlet and singlet $Z_A$, $Z_P$ and $Z_T$ renormalization constants for the $N_f=2+1+1$ ensemble cB211.072.64. 
The mass dependence is mild and we extrapolate to the chiral limit using the results at the five values of the twisted mass parameters. 

The values of the renormalization functions are listed in Table~\ref{tab:renorm}. The $N_f=2$ renormalization functions were computed in Refs.~\cite{Alexandrou:2017hac,Alexandrou:2017qyt} and are included here for easy reference. We estimate the systematic error by varying the
fit  ranges used for the extrapolation of the ${\rm RI^\prime_{MOM}}$ scale $\mu_0\to 0$. 
While the renormalization function $Z_A$ for the axial current is scheme and scale independent, the corresponding ones for the scalar and tensor charges, $Z_P$ and $Z_T$, are scale and scheme-dependent and are given  in the ${\overline{\rm MS}}$ scheme at 2~GeV. The singlet and non-singlet renormalization functions are different only for $Z_A$. For  $Z_A^{\rm s}$ we use the
conversion factor calculated to 2-loops in perturbation
theory~\cite{Skouroupathis:2008mf}. The conversion factor
 for $Z_S^{\rm s}$ and $Z_T^{\rm s}$ is
 the same as in the corresponding non-singlet case.

\begin{table}[h!]
	\centering
	{\small
		\renewcommand{\arraystretch}{1.2}
		\renewcommand{\tabcolsep}{1.5pt}
		\begin{tabular}{lc|cccccc}
			\toprule
			&&$Z_A^{ns}$&$Z_A^{s}$&$Z_P^{ns}$&$Z_P^{s}$&$Z_T^{ns}$&$Z_T^{s}$\\\hline
			$N_f$=2 &&0.7910(6)&0.797(9)&0.50(3)&0.50(2)&0.855(2)&0.852(5)\\
			$N_f$=4 &&0.763(1)& 0.753(5)&0.462(4)& 0.461(5)&0.847(1)&0.846(1)\\
			\botrule
		\end{tabular}
	}
	\caption{Non-singlet (${Z}^{ns}$) and singlet (${Z}^{s}$) renormalization constants computed using $N_f=2$  and  $N_f=4$ ensembles  and used for the renormalization of the matrix elements computed for the  cA2.09.48 and cA2.09.64 and  cB211.072.64 ensembles, respectively. We note that the scalar matrix elements in the twisted basis are renormalized with the pseudoscalar renormalization function $Z_P$. }
	\label{tab:renorm}
	\vspace*{-0.2cm}
\end{table}

\section{Results}
\label{sec:results}
\subsection{Nucleon charges}
In Table~\ref{tab:isovector}
we present our final renormalized values for the isovector charges for the three ensembles. 
Comparing the values extracted from the two $N_f=2$ ensembles with $L m_\pi\sim3$ and $L m_\pi\sim 4$   no volume effects can be resolved
within our statistical accuracy.
This corroborates our previous results at heavier than physical pion masses where no volume effects were detected for $g_A^{u-d}$~\cite{Alexandrou:2010hf}.  A previous study of cut-off effects using three $N_f=2+1+1$  ensembles with lattice spacings $a=0.089(5)$, 0.070(4), and 0.056(4)~fm, revealed that cut-off effects are negligible for a range of pion masses spanning 260~MeV to 450~MeV~\cite{Alexandrou:2010hf} for our twisted mass action. Having a clover term we expect cut-off effects to be reduced and be  within our current accuracy. However, we are planing to repeat the analysis for two further ensembles with smaller lattice spacings that will enable us to take the proper continuum limit at the physical point.

\begin{table}
	\centering
	{\small
		\renewcommand{\arraystretch}{1.2}
		\renewcommand{\tabcolsep}{3pt}
		
		\begin{tabular}{l|ccc}
			\toprule
			& $g_A^{u-d}$ & $g_S^{u-d}$ & $g_T^{u-d}$ \\ \hline
			cA2.09.48    & 1.258(27)  &  1.27(19)  &  0.954(35)  \\
			cA2.09.64    & 1.268(36)   & 0.99(28)     & 0.974(33)    \\
			cB211.072.64 & 1.286(23)  &  1.35(17)    & 0.939(27) \\
			\botrule
		\end{tabular}
	}
	\caption{Isovector charges extracted from the analysis of  the three ensembles of Table~\ref{tab:params}.}
	\label{tab:isovector}
	\vspace*{-0.2cm}
\end{table}

\begin{table}
	{
		\centering
		\small
		\renewcommand{\arraystretch}{1.2}
		\renewcommand{\tabcolsep}{3pt}
		\begin{tabular}{l|cccc}
			\toprule
			&      ${u}$       &             ${d}$              &                    ${s}$                    &                   ${c}$      \\ \hline
			$g_A$            &   0.817(29)     &            -0.450(29)           &                 -0.061(17)                 &                -0.0065(51)                 \\
			$g_S$            &    5.78(32)     &           4.51(32)            &                  0.371(38)                  &                0.059(18)                 \\
			$g_T$            &    0.737(23)     &           -0.217(23)           &                -0.0041(12)                 &                -0.0060(37)                \\
			\botrule
		\end{tabular}
	}
	
	\caption{Single flavor charges using the  cA2.09.48 ensemble.}
	\label{tab:charges_cA2}
	\vspace*{-0.2cm}
\end{table} 

\begin{figure*}
	\includegraphics[width=\linewidth]{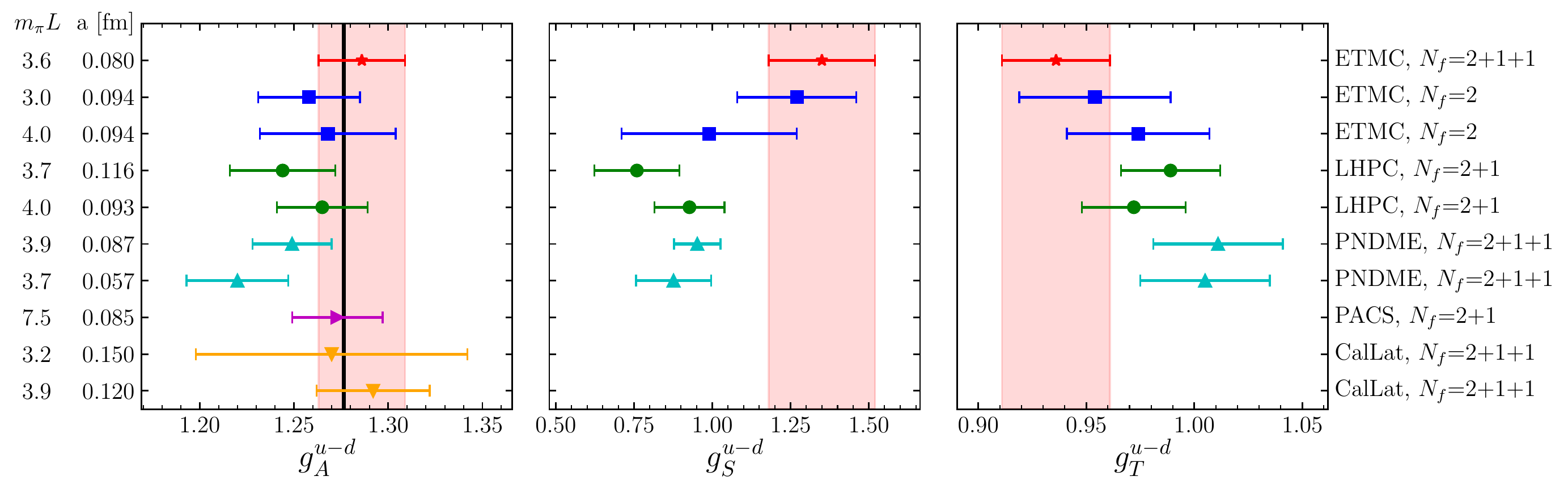}
	\vspace*{-0.8cm}
	\caption{\label{fig:comparison} Comparison of lattice QCD results on th e isovector charges computed directly at the physical point. Results are shown with a  red star for  cB211.072.64  and blue filled squares for  cA2.09.48 and cA2.09.64, green filled circles for LHPC~\cite{Hasan:2019noy}, magenta right-pointing triangle from PACS~\cite{Shintani:2018ozy}, cyan upwards-pointing triangles from PNDME~\cite{Gupta:2018qil} and  yellow downwards-pointing triangles from CalLat~\cite{Chang:2018uxx}. The line shown in the left panel is the experimental value of the nucleon axial charge $g_A^{u-d}=1.27641(56)$~\cite{Markisch:2018ndu}.}
\end{figure*}

\begin{figure}
	\includegraphics[width=0.92\linewidth]{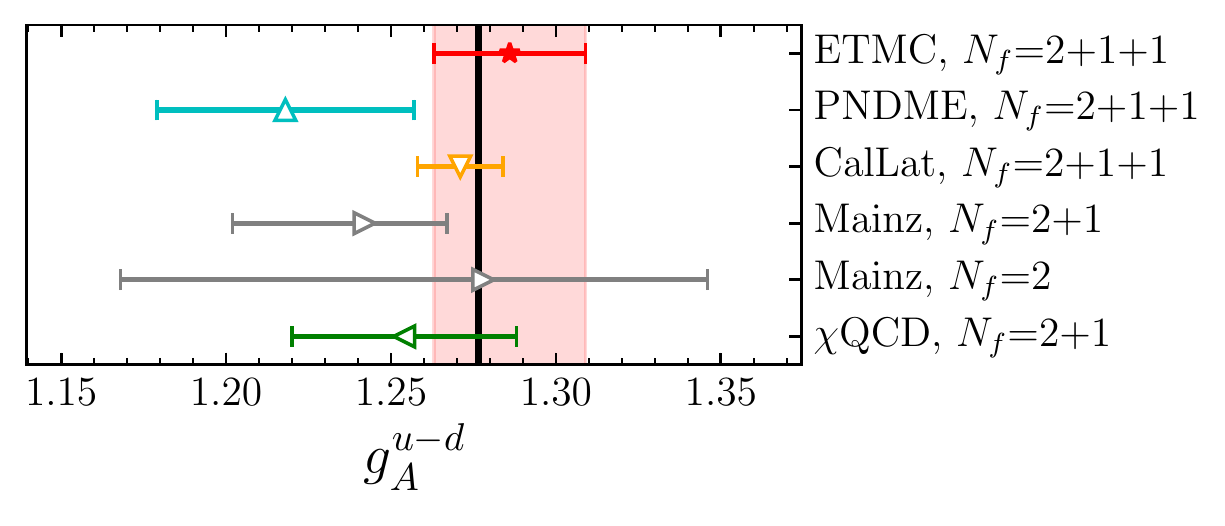}
	\vspace*{-0.1cm}
	\includegraphics[width=0.92\linewidth]{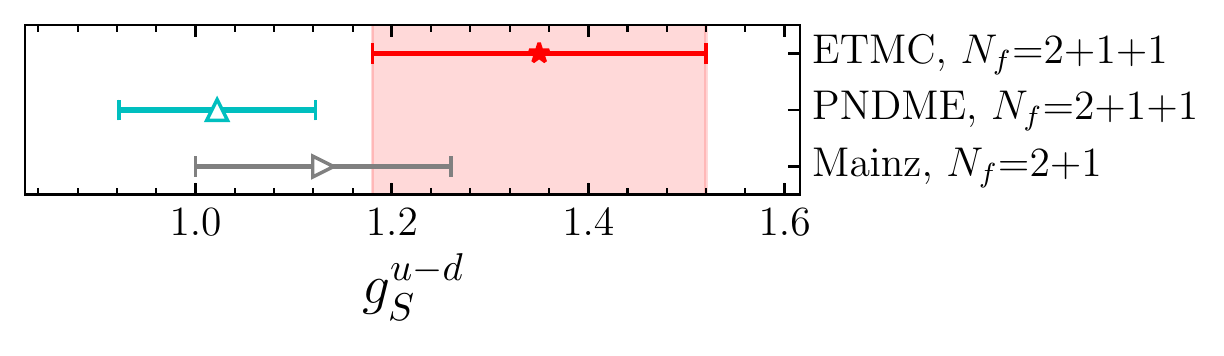}
	\vspace*{-0.1cm}
	\includegraphics[width=0.92\linewidth]{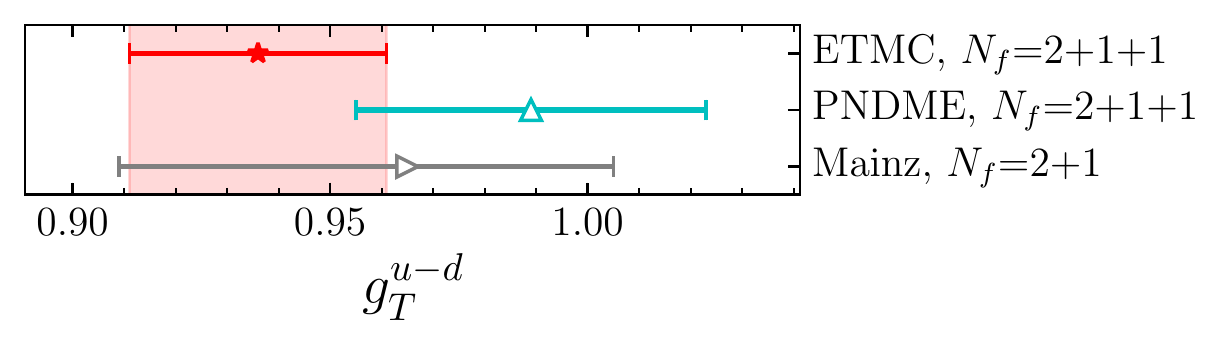}
	%\vspace*{-0.8cm}
	\caption{\label{fig:comparison_chipt} Comparison of recent lattice QCD results on the isovector charges  after chiral and continuum extrapolation (open symbols) with our results for the cB211.072.64 ensemble. Results are shown with cyan upwards-pointing triangles from PNDME~\cite{Gupta:2018qil},  yellow downwards-pointing triangles from CalLat~\cite{Chang:2018uxx},  gray right-pointing triangles from Mainz~\cite{Harris:2019bih,Capitani:2017qpc} and green left-pointing triangle from $\chi$QCD~\cite{Liang:2018pis}.  The solid line shown in the top panel is the experimental value of the nucleon axial charge $g_A^{u-d}=1.27641(56)$~\cite{Markisch:2018ndu}.}
\end{figure}

In Fig.~\ref{fig:comparison} we compare with recent results from other lattice collaborations considering only results computed using simulations with approximately physical pion mass i.e. excluding chiral extrapolations. This provides a fair comparison among the lattice QCD results. 
In Fig.~\ref{fig:comparison_chipt} we compare our results  with recent lattice QCD results obtained after performing chiral and continuum extrapolation. They include at least one ensemble with mass about 200~MeV or lower.
The final results by the  PNDME~\cite{Gupta:2018qil} and CalLat~\cite{Chang:2018uxx} collaborations shown in Fig.~\ref{fig:comparison_chipt} are obtained by  combining measurements from several ensembles with different  lattice spacing, volume and pion mass. They include  two measurements  at the physical pion mass that are also included  in Fig.~\ref{fig:comparison} as well as  the final value after a combined chiral and continuum extrapolation.  CalLat quotes as their final value $g_A^{u-d}=1.271(13)$ and PNDME  $g_A^{u-d}=1.218(25)(30)$. As can be seen, the final values by both PNDME and CalLat are in agreement with their results using the physical point ensembles, but with a much smaller error for the latter. The CalLat value is in perfect agreement with our value. Furthermore, the lattice results computed for a given ensemble over a range of lattice spacings shown in Fig.~\ref{fig:comparison} are in good agreement demonstrating that lattice spacing effects are indeed small.

For the case of $g_S^{u-d}$, we find a  value that is larger as compared to other lattice QCD determinations, which  can be explained by the fact that  $g_S^{u-d}$ increases with $t_s$. In our analysis  of the cB211.072.64 ensemble  seven values of $t_s$ are used reaching larger time separations  combined with increased statistics that allow for a better control of excited states~\cite{vonHippel:2016wid}, as demonstrated in the Appendix. Similarly, our value for  $g_T^{u-d}$ tends to be smaller since this quantity decreases with increasing values of $t_s$. As we already stressed, given that the analysis for the  cB211.072.64 ensemble is the most thorough having the largest statistics and the biggest number of $t_s$, we consider as final the values extracted using this ensemble.
In Fig.~\ref{fig:comparison_chipt} we include the values of $g_S^{u-d}$ and $g_T^{u-d}$ obtained by the PNDME and CLS collaborations after chiral and continuum extrapolation. The chirally extrapolated values by PNDME are consistent with their value using the two physical ensembles, corroborating the fact that finite discretization effects are small and consistent with our findings using heavier than physical pion mass~\cite{Alexandrou:2010hf}. The computation by the CLS Mainz group used ensembles with a smallest pion mass of about 200~MeV and it is in agreement with our values.   

\begin{table}
	{
		\centering
		\small
		\renewcommand{\arraystretch}{1.2}
		\renewcommand{\tabcolsep}{3pt}
		\begin{tabular}{l|cccc}
			\toprule
			&      ${u\texttt{-}d}$       &             ${u\texttt{+}d\texttt{-}2s}$              &                    ${u\texttt{+}d\texttt{+}s\texttt{-}3c}$                    &                   ${u\texttt{+}d\texttt{+}s\texttt{+}c}$      \\ \hline
			$g_A$            &    1.286(23)     &           0.530(18)           &                 0.422(25)                 &                0.382(31)                 \\
			$g_S$            &    1.35(17)     &             9.92(90)            &                  11.1(1.0)                  &                11.4(1.0)                  \\
			$g_T$            &  0.936(25)     &           0.527(22)           &                0.519(22)                 &                  0.518(22)                \\
			\botrule
			%		\toprule
			&      ${u}$       &             ${d}$              &                    ${s}$                    &                   ${c}$      \\ \hline
			$g_A$            &    0.862(17)     &            -0.424(16)           &                 -0.0458(73)                 &                -0.0098(34)                 \\
			$g_S$            &    6.09(55)     &           4.74(43)            &                  0.454(61)                  &                0.075(17)                 \\
			$g_T$            &    0.729(22)     &           -0.2075(75)           &                -0.00268(58)                 &                 -0.00024(16)                \\
			\botrule
		\end{tabular}
	}
	
	\caption{Isovector, isoscalar and single flavor charges using the  cB211.072.64 ensemble.}
	\label{tab:charges}
	\vspace*{-0.2cm}
\end{table} 

The values extracted for the renormalized isovector, isoscalar and single flavor charges   are tabulated in Tables~\ref{tab:charges_cA2} and~\ref{tab:charges} for the cA2.09.48  and cB211.072.64 ensembles respectively. The latter are our best determination of these quantities and in particular the precision obtained in the determination of  the single flavor charges  computed directly at the physical point using this ensemble is much better as compared to any other available lattice QCD results. This includes also our previous determination~\cite{Alexandrou:2017qyt,Alexandrou:2017hac} using  the cA2.09.48 ensemble. In particular, we find for the first time, for $g_A^c$, $g_S^c$ and $g_T^c$  a non-zero value showing charm quark effects.

\begin{figure}
	\centering
	\hspace*{-0.2cm}\includegraphics[width=0.9\linewidth]{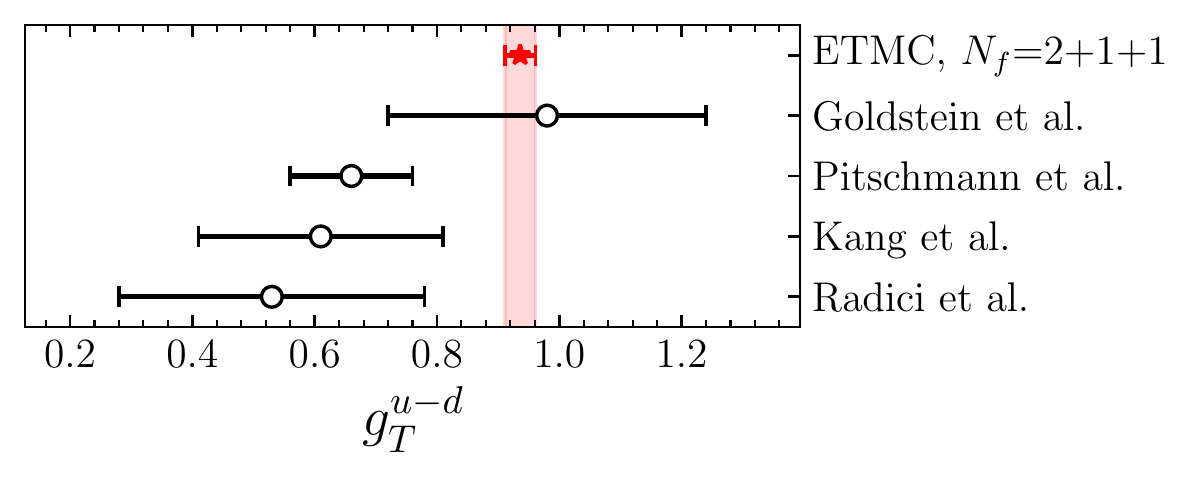}\\
	\vspace*{-0.1cm}
	\includegraphics[width=0.92\linewidth]{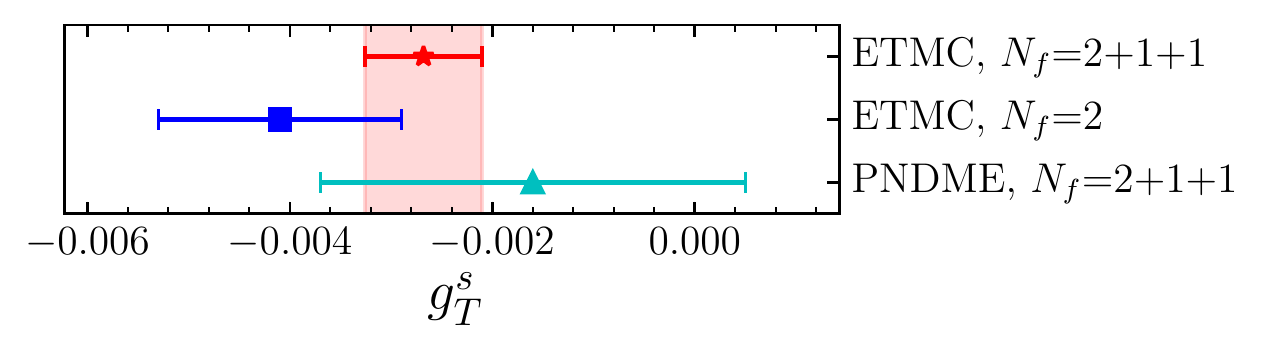}\\
	\vspace*{-0.3cm}
	\caption{\label{fig:comparison_gT} The isovector (top) and strange (bottom) nucleon tensor charges computed directly at the physical point by various lattice QCD collaborations. We include phenomenological results when available. We exclude phenomenological results that use input from lattice QCD. Results are shown with a  red star for the  cB211.072.64  ensemble, with blue filled squares for the  cA2.09.48 ensemble and  with cyan upwards-pointing triangles from PNDME~\cite{Gupta:2018lvp}. 
	Phenomenological results are shown with open black circles~\cite{Goldstein:2014aja,Pitschmann:2014jxa,Kang:2015msa,Radici:2018iag}. For $g_T^{u-d}$ we only compare with the value extracted using the cB211.072.64 ensemble since other lattice results are given in Fig.~\ref{fig:comparison}.}
	%\vspace*{-0.4cm}
\end{figure}

\begin{figure*}
	\includegraphics[width=\linewidth]{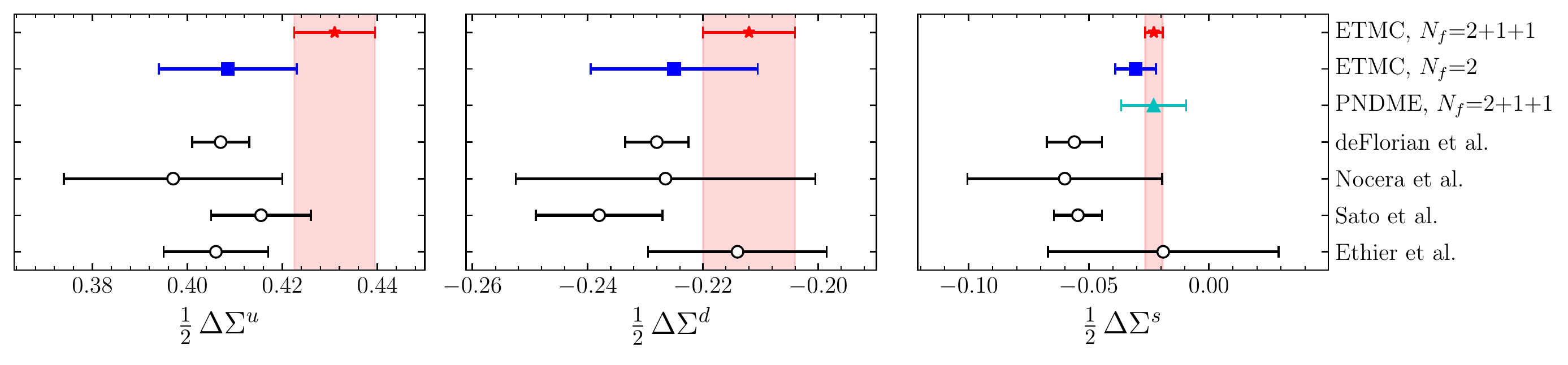}\\
	\vspace*{-0.3cm}
	\caption{\label{fig:comparison_gA} The intrinsic spin $\frac{1}{2}\,\Delta\Sigma^q$ carried by the  up, down and strange quarks in the nucleon computed directly at the physical point by various lattice QCD collaborations. We include results from  phenomenological analyses when available. Lattice QCD results are shown with a  red star for  the cB211.072.64, with blue filled squares for the  cA2.09.48 ensemble and with cyan upwards-pointing triangles from PNDME~\cite{Lin:2018obj}. Phenomenological results are shown with open black circles~\cite{deFlorian:2009vb,Nocera:2014gqa,Sato:2016tuz,Ethier:2017zbq}.}
	%\vspace*{-0.4cm}
\end{figure*}

In Figs.~\ref{fig:comparison_gT} and~\ref{fig:comparison_gA} we show a comparison of available lattice QCD results for the tensor charges and intrinsic spin contributions of quarks, $1/2\Delta\Sigma^{u,d,s}$, to the proton computed directly at the physical point\footnote{The results of the PNDME collaboration for the connected contributions and for the strange charges are of comparable quality~\cite{Gupta:2018qil,Lin:2018obj}. We note that the result of PNDME after chiral and continuum extrapolation is consistent with their value using the physical ensemble. However, the  disconnected contributions to the up and down quarks have not been computed at the physical point and thus we do not include them in Fig.~\ref{fig:comparison_gA}. One can find the values without the disconnected contributions in Ref.~\cite{Lin:2018obj}.}.   We also include phenomenological results, where we limit ourselves to  those that  have not used input from lattice QCD. A good agreement is observed among lattice QCD results and a notable observation  is that the lattice QCD results are at least as accurate as the phenomenological determinations.

 PCAC relates  $g_A^{u-d}$ to the pseudoscalar charge through the relation $g_P^{u-d} = m_N/m_{ud}\, g_A^{u-d}$~\cite{Gonzalez-Alonso:2013ura}. Using our values for $g_A^{u-d}$, the relation
\begin{equation}\label{eq:mud}
a\,m_{ud}\,Z_P = a\,\mu = 0.00072,
\end{equation}
where $a \mu$ is the twisted mass parameter in lattice units used in the simulation of the cB211.072.64 ensemble,
and the extracted value of the nucleon mass  $m_N$ we obtain $g_P^{u-d} =313.8(6.4)$. This value is lower as compared to $g_P^{u-d} =349(9)$ found  in Ref.~\cite{Gonzalez-Alonso:2013ura}. A direct evaluation of $g_P^{u-d}$ in lattice QCD using the same setup will be undertaken in the future to study the origin of this discrepancy.

\subsection{Nucleon $\sigma$-terms}
\begin{table}
	{
		\centering
		\small
		\renewcommand{\arraystretch}{1.2}
		\renewcommand{\tabcolsep}{3pt}
		\begin{tabular}{l|ccc}
			\toprule
			               & ${u+d}$ &   ${s}$    &   ${c}$   \\ \hline
			$\sigma$ [MeV] &     41.6(3.8)     & 45.6(6.2)  &  107(22)  \\
			$f^N$          &    0.0444(43)     & 0.0487(68) & 0.115(24)\\
			\botrule  
		\end{tabular}
	}
	
	\caption{Isoscalar and single flavor $\sigma$-terms and $f^N$ using the  cB211.072.64 ensemble. Instead of giving the $\sigma$-term for the up and down quarks separately we give the isoscalar combination which is the  phenomenologically relevant  $\sigma_{\pi N}$. We also give the factors $f^{N}_{ud} \equiv \sigma_{\pi N}/m_N$. }
	\label{tab:sigma}
	\vspace*{-0.2cm}
\end{table} 
\begin{figure}
	\includegraphics[width=0.92\linewidth]{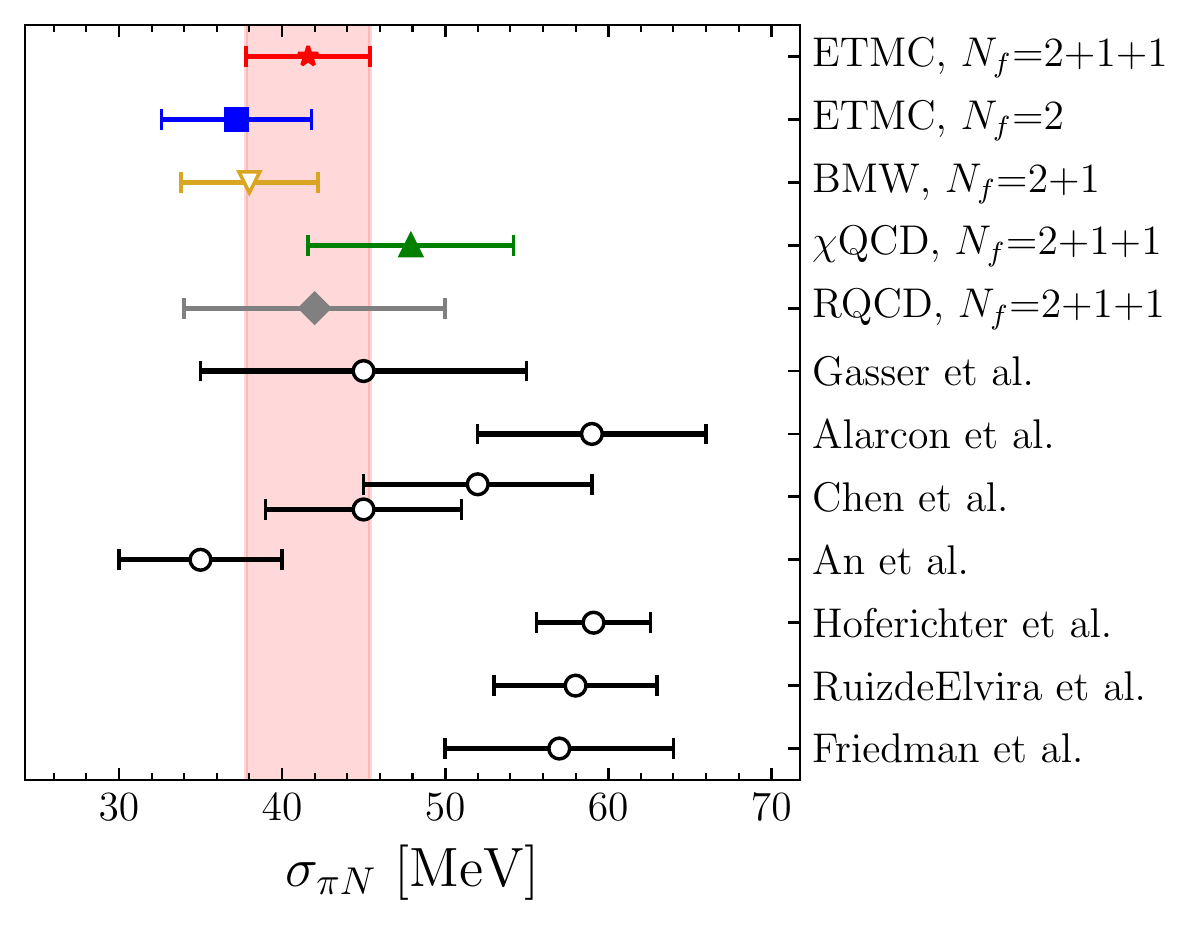}\\
	\includegraphics[width=0.92\linewidth]{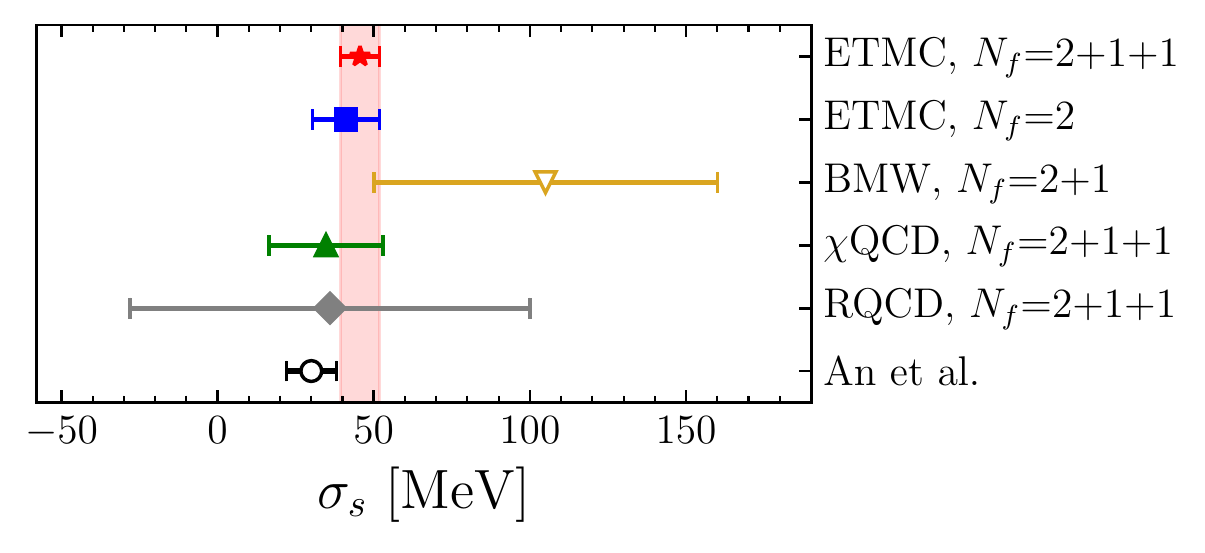}\\
	\vspace*{-0.3cm}
	\caption{\label{fig:comparison_sigma} Nucleon $\sigma$-terms from lattice QCD and from phenomenology. We show lattice QCD results computed directly at the physical point by evaluating the three-point functions with filled symbols, namely for the cB211.072.64 ensemble (red star) for the cA2.09.48 ensembles~\cite{Abdel-Rehim:2016won} (blue square), from $\chi$QCD~\cite{Yang:2015uis} (green up-pointing triangles), and from RQCD~\cite{Bali:2016lvx} (gray diamond). We also show a result form BMW~\cite{Durr:2015dna} obtained using the Feynman-Helmann method that includes ensembles at the physical point (open down-pointing triangle). We note that the results by RQCD and $\chi$QCD  after chiral and continuum extrapolation agree with their corresponding value obtained using their physical point ensemble and are thus not included.  We include a range of phenomenological results that do not use lattice QCD input~\cite{Gasser:1990ce,Alarcon:2011zs,Chen:2012nx,An:2014aea,Hoferichter:2015dsa,RuizdeElvira:2017stg,Friedman:2019zhc} (open circles).}
	%\vspace*{-0.4cm}
\end{figure}
The  nucleon $\sigma$-terms that give the scalar quark contents  are fundamental quantities of QCD. They
determine the mass generated by the quarks in the nucleon. They are relevant
for a wide range of physical processes and for the interpretation of direct-detection dark
matter (DM) searches~\cite{Giedt:2009mr} being undertaken by a number of
experiments~\cite{Cushman:2013zza}.  It is
customary to define the nucleon $\sigma$-terms to be scheme- and
scale-independent quantities:
\begin{equation}
\sigma_f=m_{q_f} \langle N|\bar{q}_f q_f|N\rangle ,\,\,
\sigma_{\pi N}= m_{ud}\langle N | \bar{u}u + \bar{d}d | N \rangle
\end{equation}
for a given quark $q_f$ of flavor $f$,  or for the isoscalar
combination, where $m_{q_f}$ is the mass of $q_f$,
$m_{ud}=(m_u+m_d)/2$ is the average light quark mass and $|N\rangle$
is the nucleon state.

Since the pioneering chiral perturbation theory analysis that yielded
$\sigma_{\pi N}\sim 45$~MeV~\cite{Gasser:1990ce}, there has been significant progress in the
determination of $\sigma_{\pi N}$ from experimental data~\cite{Hoferichter:2012wf,Alarcon:2011zs}. Using
high-precision data from pionic atoms to determine the $\pi
N$-scattering lengths and a system of Roy-Steiner equations that
encode constraints from analyticity, unitarity, and crossing symmetry
a value of $59.1(3.5)$~MeV is obtained~\cite{Hoferichter:2015dsa}.
% A more recent chiral perturbation theory analysis using new
%pion-nucleon scattering data, and a precise determination of the
%$\pi-N$ scattering lengths from pionic-atoms yields a value of
%$59(7)$~MeV~\cite{Alarcon:2011zs}.  An analysis using the same data
%and a system of Roy-Steiner equations finds
This larger value of $\sigma_{\pi N}$ has theoretical implications on
our understanding of the strong interactions as stressed in
Ref.~\cite{Leutwyler:2015jga}.  Given the importance of these quantities, a number of lattice QCD
calculations have been undertaken to compute them using two
approaches~\cite{Young:2009ps}. The first uses the Feynman-Hellmann
theorem that is based on the variation of the nucleon mass $m_N$ with
$m_{q_f}$:~$\sigma_{f}=m_{q_f}\frac{\partial m_N}{\partial  m_{q_f}}$. However, since the dependence of the nucleon mass on the strange and charm quark mass is weak, this approach yields large errors. An alternative method is to evaluate directly the nucleon matrix
elements of the scalar operator that involves 
disconnected quark loops as done in this work. The evaluation of the three-point function is computationally  much more demanding 
than hadron masses. Therefore, it is only recently that
a direct computation of the $\sigma$-terms has been performed using
dynamical simulations~\cite{Bali:2011ks,Freeman:2012ry,Gong:2013vja,Abdel-Rehim:2013wlz,Alexandrou:2013nda,Yang:2015uis, Bali:2016lvx}.

The values extracted for the $\sigma$-terms and $f^N$ are tabulated in Table~\ref{tab:sigma} for the cB211.072.64 ensemble.
We show results for the $\sigma_{\pi N}$- and $\sigma_{s}$-terms in Fig.~\ref{fig:comparison_sigma}
computed by various lattice QCD collaborations at the physical point either directly by computing the three-point function or using the Feynman-Hellmann method that include ensembles at the physical point. We also compare with phenomenological results limiting ourselves to those that have not used input from lattice QCD. We observe a very good agreement among lattice QCD results. One can see the very large uncertainty on  $\sigma_{s}$ by BMW when extracted using the Feynman-Hellmann method.

It is customary to also provide  results in terms of the
dimensionless ratios, $f^N_f = \sigma_f/m_N$. Since we have the isovector matrix element $\langle N|\bar{u}u -
\bar{d}d|N\rangle$, for the cB211.072.64 ensemble, we can combine it with the isoscalar
matrix element to obtain the individual up- and down-quark
contributions for the proton and the neutron in the isospin limit via the relations
\begin{align}
  f_u^p = \frac{2 m_{ud} r}{r+1} \frac{\langle N|\bar{u}u|N\rangle}{m_N} & & f_u^n = \frac{2 m_{ud} r}{r+1} \frac{\langle N|\bar{d}d|N\rangle}{m_N} \nonumber\\
  f_d^p = \frac{2 m_{ud} }{r+1} \frac{\langle N|\bar{d}d|N\rangle}{m_N} & & f_d^n = \frac{2 m_{ud}}{r+1} \frac{\langle N|\bar{u}u|N\rangle}{m_N},
\end{align}
where the up and down quark mass splitting entering in $f_u^N$ and $f_d^N$ is computed taking the ratio of  the up to the down quark masses
%$\mu = \frac{m_u+m_d}{2}$ is the twisted quark mass and
$r=m_u/m_d=0.513(31)$ from
Ref.~\cite{Aoki:2019cca} together with our determination of the up and down quark mass, as given in  Eq.~\eqref{eq:mud}. We obtain
\begin{align}
	f_u^p = 0.0169(18),  &  & f_u^n = 0.0132(14), \nonumber \\
	f_d^p =  0.0257(26), &  & f_d^n =0.0330(33).
\end{align}
These results are compatible with the results of our previous study using the cA2.09.48 ensemble~\cite{Abdel-Rehim:2016won}.
The isovector scalar charge is also related to the neutron-proton mass splitting $\delta m_N^{\rm QCD}$ in the absence of electromagnetism~\cite{Gonzalez-Alonso:2013ura} through the relation $\langle N | \bar{u}u - \bar{d}d | N
\rangle = \Delta m_N^{\rm QCD}/\Delta m_{ud}$. We thus obtain
\begin{equation}
\Delta m_N^{\rm QCD} = 2 m_{ud} \frac{1-r}{1+r}\langle N | \bar{u}u - \bar{d}d | N
\rangle
\end{equation}
and we find  $\Delta m_N^{\rm QCD} = 3.33(50)$~MeV.
This value is consistent with   $\Delta m_N^{\rm QCD} = 2.52(17)(24)$~MeV  determined for
non-degenerate up- and down-quarks~\cite{Borsanyi:2014jba}.  

The y-parameter, defined as $y=2\frac{\langle N|\bar{s}s|N \rangle}{\langle N|\bar{u}u+\bar{d}d|N\rangle}$, gives a measure of the  strangeness content of the nucleon. We find a value of $y =0.0849(81)$ for the cB211.072.64 ensemble.

\section{Conclusions}
\label{sec:conclusions}
 Results on the nucleon axial, tensor and scalar charges are presented for three ensembles of twisted mass clover-improved fermions  tuned to reproduce the  physical value of the pion mass. The most thorough analysis is performed for the $N_f=2+1+1$ ensemble which provides one of the best descriptions of the QCD vacuum to date having  light, strange and charm quarks in the sea.
A notable result of this work is the accurate computation of $g_A^{u-d}$ using the $N_f=2+1+1$ cB211.072.64 ensemble that  agrees with the experimental value of 1.27641(56)~\cite{Markisch:2018ndu}. An additional milestone is the evaluation to an unprecedented accuracy of the flavor charges directly at the physical point taking into account the disconnected contributions. We show that the charm axial  charge is non-zero and obtain  a value for $g_A^s$ that is more  accurate than recent phenomenological determinations. It confirms the smaller values recently  suggested by the NNPDF~\cite{Nocera:2014gqa}  and JAM17~\cite{Ethier:2017zbq} analyses both of which, however, carry a large error.
We find  that the intrinsic quark spin contribution in the nucleon is  $\frac{1}{2}\Delta \Sigma=\frac{1}{2}\sum_{f=u,d,s,c} g_A^f=0.191(16)$. The non-singlet combination is found to be $g_A^{u+d-2s}=0.530(18)$.
Furthermore, the  evaluation of the isovector scalar and tensor charges to an accuracy of about 10\% and 3\%,  respectively provides valuable input to experimental studies  on  possible allowed scalar and tensor interactions and new physics searches~\cite{Bhattacharya:2011qm}.

Using the scalar matrix element we extract the nucleon $\sigma$-terms that are important for direct dark matter searches and for phenomenological studies  of $\pi N$ scattering processes. We find $\sigma_{\pi N}=41.6(3.8)$~MeV, that confirms a smaller value already suggested from previous lattice QCD studies~\cite{Abdel-Rehim:2016won,Yang:2015uis,Bali:2011ks}. While this smaller value is in agreement with the first analysis that yielded
$\sigma_{\pi N}\sim 45$~MeV~\cite{Gasser:1990ce},  it is in tension 
with recent analyses that yield larger values. An analysis based on the Roy-Steiner equations
and experimental data on pionic atoms extracted the  value 
of $59.1(3.5)$~MeV~\cite{Hoferichter:2015dsa} that is confirmed by using a large-scale fit of pionic-atom level shift and width data across the periodic table~\cite{Friedman:2019zhc}. The larger value is also confirmed by  using the $\pi N$  scattering lengths from the low-energy data base~\cite{RuizdeElvira:2017stg}.  Given the significant progress in the
determination of $\sigma_{\pi N}$ both using experimental data~\cite{Hoferichter:2015hva,Hoferichter:2012wf,Alarcon:2011zs} and lattice QCD this persisting tension needs to be further examined. Computing the $\pi N$ scattering lengths within lattice QCD will provide a crucial cross-check.

\section*{ACKNOWLEDGMENTS}
We acknowledge funding from the European Union's Horizon 2020 research and innovation programme under the Marie Sklodowska-Curie grant agreement No 642069 and from the COMPLEMENTARY/0916/0015 project funded by the Cyprus Research Promotion Foundation. M.C. acknowledges financial support by the U.S. National Science
Foundation under Grant No.~PHY-1714407. This work was supported by a grant from the Swiss National Supercomputing Centre (CSCS) under project ID s702. We thank the staff of CSCS for access to the computational resources and for their constant support. The Gauss Centre for Supercomputing e.V. (www.gauss-centre.eu) funded
the project pr74yo by providing computing time on the GCS Supercomputer SuperMUC at Leibniz Supercomputing Centre (www.lrz.de).  In addition, this work used computational resources from Extreme Science and
Engineering Discovery Environment (XSEDE),
which is supported by National Science Foundation grant number TG-PHY170022.
This work used computational resources from the John von Neumann-Institute for Computing on the JUWELS system at the research center in Jülich, under the project with id ECY00 and HCH02.
K.H. is financially supported by  the  Cyprus  Research  Promotion  foundation  under contract  number  POST-DOC/0718/0100.

\section*{APPENDIX}

We examine here in more detail the importance of increasing the statistics in the three-point functions as we  increase the source-sink time separation in order to keep the statistical error approximately constant. This is carried out for the cB211.072.64 ensemble for which we increase the statistics as  listed in Table~\ref{tab:statistics} at each $t_s$. As we will show below, excited state effects are only correctly addressed if the statistics are sufficiently large to keep the errors approximately the same among the various values of $t_s$. 

\begin{figure}
	\includegraphics[width=0.9\linewidth]{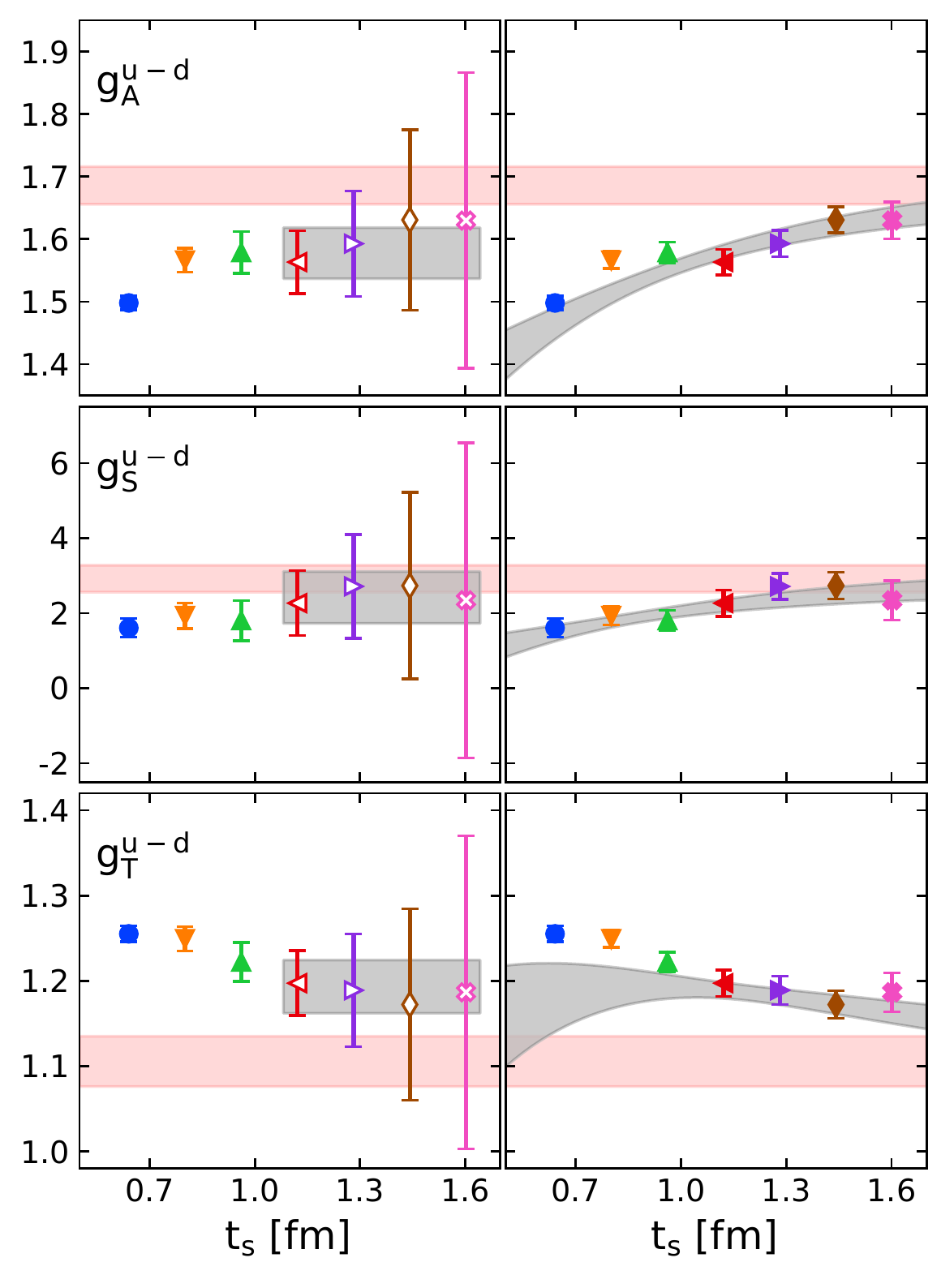}\\
	\vspace*{-0.3cm}
	\caption{\label{fig:plateau_wrong} We show the plateau values of the ratio that yields the  bare isovector charges for the cB211.072.64 ensemble. The  left panels show the plateau values   obtained keeping the statistics the same as those of the smallest $t_s$. The open symbols show the plateau values one would consider converged given the errors. The gray band is the weighted average of the  plateau values for $t_s \stackrel{>}{=} 1.2$~fm.
          The right panels show the plateau values  when increasing the statistics
          with increasing  source-sink time separation as listed in Table~\ref{tab:statistics}. The grey  bands are the same as those shown in Fig.~\ref{fig:cB211.64_gA} as are the red bands across both left and right panels.}
	\vspace*{-0.4cm}
\end{figure}

\begin{figure*}
	\includegraphics[width=\linewidth]{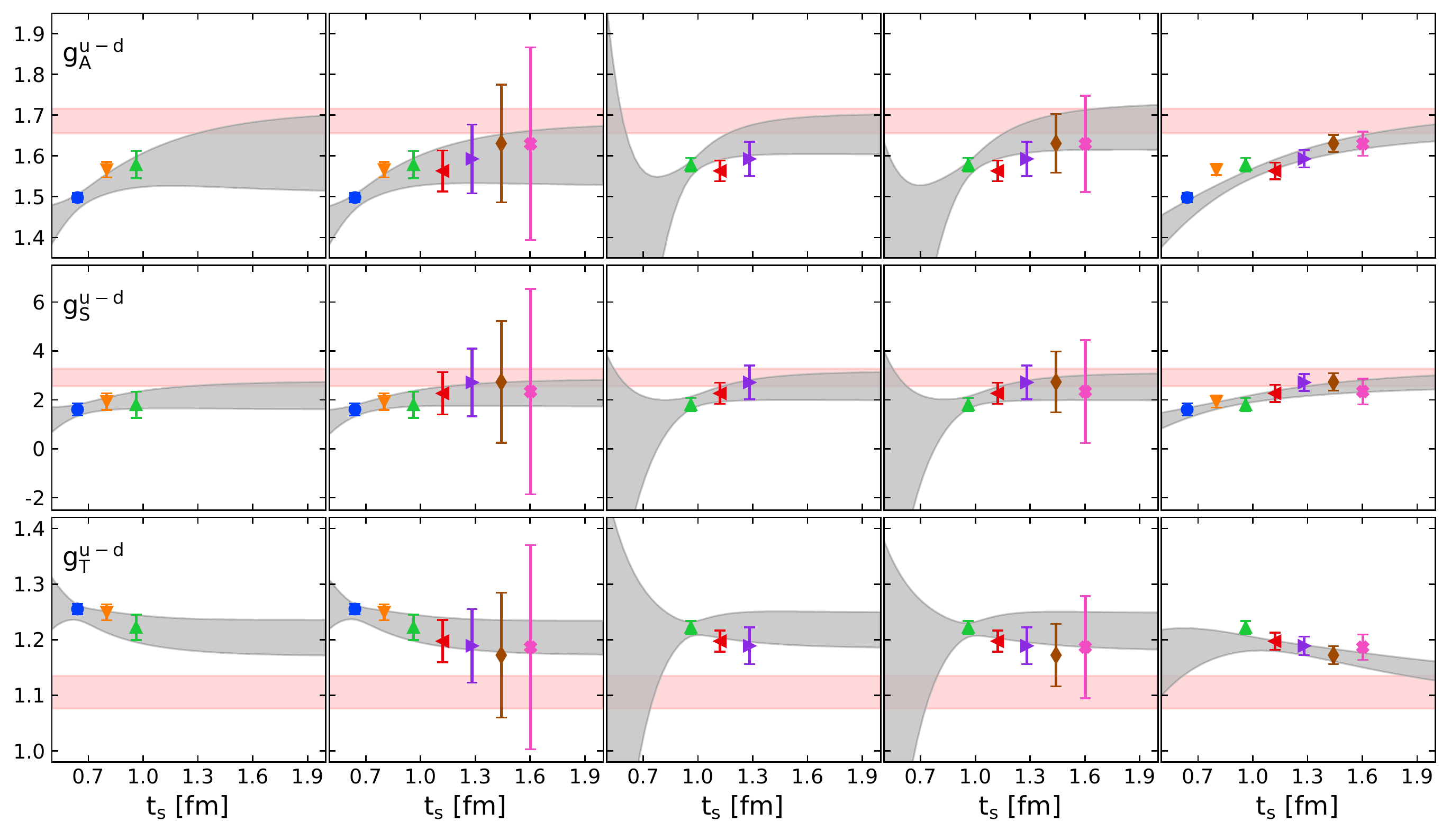}\\
	\vspace*{-0.3cm}
	\caption{\label{fig:two_state_wrong} We show the values extracted for the bare isovector  charges for the cB211.072.64 ensemble using the two state fit approach. The notation is the same as that in Fig~\ref{fig:two-state} showing in each panel  the plateau values at the $t_s$ values included in the two-state fits. The  two leftmost panels show the resulting fit setting $t_{\rm ins}=t_s/2$ as a function of $t_s$ (grey bands) when 750 measurements are used for each $t_s$, i.e. the statistics used for the smallest $t_s=0.64$~fm. The third and fourth two panels show the resulting fits when using  3000 measurements for each $t_s$, i.e. the statistics used for $t_s=0.96$~fm. The rightmost panel shows the results using the full statistics, from which the final values for the bare charges are obtained as depicted by the red band across all panels.}
	\vspace*{-0.4cm}
\end{figure*}

\subsection{Plateau method}
\label{sec:plateau}

We first examine the plateau method for the case of the isovector charge operators.
We show in the left panel of Fig.~\ref{fig:plateau_wrong} the plateau values that one  would obtain if one kept  the statistics the same as at the smallest $t_s$ value  for all the source-sink time separations instead of the ones given in  Table~\ref{tab:statistics}. As can be seen the errors increase becoming very large for the two largest time separations. With such errors one might think that convergence is reached already at $t_s=1.2$~fm, shown by the open red symbol. For $g_S^{u-d}$ for which effects of excited state are milder one would obtain a value compatible with the one extracted from the two-state fit. However, a weighted average would yield a lower value for $g_A^{u-d}$ and a larger one for $g_T^{u-d}$, as shown by the gray bands. This is to be contrasted with the values extracted
using the increased statistics of Table~\ref{tab:statistics} for larger $t_s$ values, shown in the right panel of Fig.~\ref{fig:plateau_wrong}.
 For both $g_A^{u-d}$ and $g_T^{u-d}$ there is a clear indication that excited states are still present and even larger time separations are needed to be sure that one has converged to the two-state result depicted by the red band.

\subsection{Two-state fit method}

Our data show also that the two-state fit approach cannot capture correctly excited state effects if the error increases with the source-sink separation. This is specially seen for the tensor charge as depicted
in Fig.~\ref{fig:two_state_wrong} we show results extracted using the two-state fit approach considering five different cases:
\begin{itemize}
\item Using the three smaller values of $t_s$ keeping the statistics the same as that of the smallest $t_s$, namely 750. Since the smallest $t_s$ is the most accurate its weight in the fit is large and the error band increases as $t_s$ increases, resulting in a mean value for $g_A^{u-d}$ and $g_S^{u-d}$ that is below the one obtained if one uses the full statistics shown by the red band. Given the large error in particular for $g_A^{u-d}$ the two results are consistent. However, for $g_T^{u-d}$ the fit yields a value that clearly overestimates the value extracted when using the full statistics.
\item Using all values of $t_s$ but keeping the statistics at 750 for all. A similar behavior is observed despite the fact that we now have seven time separations instead of three since the fits are dominated by the smallest and most accurate point. Therefore, having results at larger time separations, without increasing statistics is not very useful.  
\item Using $t_s=0.96$~ fm, 1.12~fm and 1.28~fm keeping the statistics at 3000, namely the same as for the smallest $t_s$ used in the fit. Having larger time separations with more statistics tends to increase the mean values of $g_A^{u-d}$ and $g_s^{u-d}$, but once more the fit is heavily biased b the first most accurate point, resulting in a larger value for $g_T^{u-d}$.
\item Using all values greater than $t_s=0.96$~fm but keeping the statistics the same as for $t_s=0.96$~fm. Similar results are obtained as with the previous case.
\item Using the full statistics as listed in Table~\ref{tab:statistics} for our final analysis. For the extraction of the final value $g_T^{u-d}$ we exclude the two smallest time separations to ensure agreement with the summation method as shown in Fig.~\ref{fig:cB211.64_gT}.
\end{itemize}
As this study shows it is important to both have sink-source time separations that span a large enough range and also to increase statistics so the errors at each time separation are approximately constant. Otherwise, the two-state fit is driven by the most accurate point and can lead to wrong results.

\bibliography{refs}

\end{document}